\documentclass[letterpaper]{article} 
\usepackage{aaai25}  
\usepackage{times}  
\usepackage{helvet}  
\usepackage{courier}  
\usepackage[hyphens]{url}  
\usepackage{graphicx} 
\usepackage{natbib}  
\usepackage{caption} 
\usepackage{amsmath}
\usepackage{amsthm}
\usepackage{multirow}
\usepackage{multicol}
\usepackage{amsfonts}
\usepackage{enumitem}
\usepackage{subfigure}
\usepackage{color}
\usepackage{booktabs}

\usepackage[linesnumbered,ruled,vlined]{algorithm2e}
\nocopyright

\title{VVRec: Reconstruction Attacks on DL-based Volumetric Video Upstreaming via Latent Diffusion Model with Gamma Distribution}
\author{
    Rui Lu\equalcontrib, 
    Bihai Zhang\equalcontrib, 
    Dan Wang
}
\affiliations{
    The Hong Kong Polytechnic University\\
    \{csrlu, csbzhang, csdwang\}@comp.polyu.edu.hk
}

\newcommand{\pointcloud}{\mathbf{x}}
\newcommand{\shapelat}{\mathbf{z}}

\newcommand{\latpoint}{\mathbf{h}}
\begin{document}

\maketitle

\begin{abstract}
With the popularity of 3D volumetric video applications, such as Autonomous Driving, Virtual Reality, and Mixed Reality, current developers have turned to deep learning for compressing volumetric video frames, i.e., point clouds for video upstreaming. The latest deep learning-based solutions offer higher efficiency, lower distortion, and better hardware support than traditional ones like MPEG and JPEG. However, privacy threats arise, especially \textit{reconstruction attacks} targeting to recover the original input point cloud from the intermediate results.
In this paper, we design \textbf{VVRec}, to the best of our knowledge, which is the first targeting DL-based \textbf{\underline{V}}olumetric \textbf{\underline{V}}ideo \textbf{\underline{Rec}}onstruction attack scheme.
VVRec demonstrates the ability to reconstruct high-quality point clouds from intercepted transmission intermediate results using four well-trained neural network modules we design. Leveraging the latest latent diffusion models with Gamma distribution and a refinement algorithm, VVRec excels in reconstruction quality, and color recovery, and surpasses existing defenses.
We evaluate VVRec using three volumetric video datasets. The results demonstrate that VVRec achieves 64.70dB reconstruction accuracy, with an impressive 46.39\% reduction of distortion over baselines.
\end{abstract}

\section{Introduction}
Currently, with the increasing broadcasting of 3D applications, customers can enjoy and experience the volumetric video technical, which are from fiction in the past, now coming to daily life using 3D commercial devices like Meta Oculus, Microsoft Hololens, etc.
Emerging 3D applications in domains such as the Metaverse, Esports, and VR/MR require users to interactively share and display their surroundings with others. Unlike previous downstream applications where users only acted as video content receivers from cloud servers, these upstream applications, such as virtual meetings and VR games, necessitate active participation and real-time interaction.
Every user is a unique content provider, and even a smartphone, e.g., the latest iPhone, can capture volumetric videos for Apple Vision Pro display and upstream them to others~\cite{apple-2023}. 

Existing volumetric video is in the format of \textit{point clouds} offering high precision and rich color representation~\cite{zhang2023gpccpp}. Simultaneously, streaming techniques of volumetric video are developing rapidly~\cite{patchdpcc} and public standards are introduced, e.g., G-PCC and V-PCC proposed by the ISO Moving Picture Experts Group (MPEG), and Google Draco and JPEG, etc. They apply a series of logical computations and contextual encoding based on octree structure features or 2D to 3D projection.

\begin{figure}[t]
    \centering
    \includegraphics[width=0.9\linewidth]{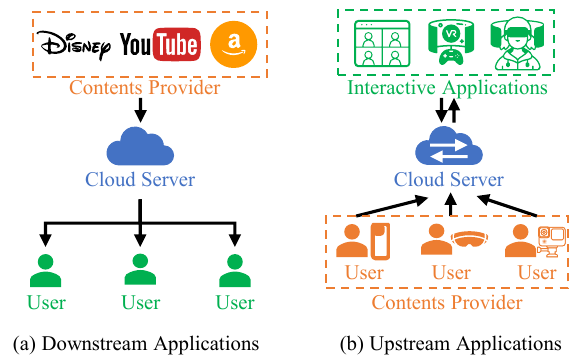}
    \caption{Video applications where the content provider is evolving, from filming companies to individual users, can be categorized into Downstream and Upstream Applications.}
    \label{fig:enter-label}
\end{figure}

Deep learning methods, known for their capability in context understanding and information encoding, have attracted attention, particularly in the volumetric video compression area~\cite{scp,gao2022openpointcloud}.
A DL-based approach involves constructing and training a pair of neural network (NN) models as an encoder and a decoder, compressing point clouds during NN inference~\cite{quach2019learning}. 
For instance, LVAC~\cite{lvac} utilizes a local coordinate-based model to input point clouds and compress them into intermediate results. These intermediate results are then transmitted to the cloud, where a corresponding decoder reverses them back to the original point clouds. In comparison to non-DL-based like G-PCC, V-PCC, DL-based methods offer advantages such as low distortion~\cite{pc_survey}, excellent compression efficiency~\cite{3dac}, and great hardware supports~\cite{octformer}.
However, \textit{privacy concerns} remain a significant threat during volumetric video serving~\cite{lu2023pagoda}. One of the attacks that can recover the original data from intermediate results is known as \textit{Reconstruction Attacks}. It occurs during NN model inference across various machine-learning domains. For example, in language models, reconstruction attacks aim to reconstruct input text~\cite{song2020eia,brown2022does}, while in computer vision, they aim to reconstruct input images~\cite{dosovitskiy2016inverting}. However, no existing research has been done on activating the reconstruction attack on the upstreaming point cloud data.

The major challenge to attack DL-based point cloud compression models is the \textit{distribution shift} between the training dataset used by the encoder-decoder pair of the service provider cloud and that used by attackers. This disparity prevents attackers from precisely training an identical decoder, resulting in significant reconstruction distortion. This disparity is attributed to the \textit{distribution shift} in the training dataset between attackers and the encoder-decoder, which has been discussed widely in many machine learning domains, e.g., Transfer Learning~\cite{zhuang2020comprehensive}, Federated Learning~\cite{reisizadeh2020robust}, etc. This shift worsens in point cloud data as it owns higher dimensional recording of the geometry information~\cite{9127813}. Another challenge is that point clouds have a dynamic scale, often downsampled during compression by NN encoders for higher compression rates. Unlike images with specific resolutions, point clouds lacking precise scale may face significant degradation in reconstruction performance. The sampling or scale prediction is necessary for the attacker.

In our paper, we design \textbf{VVRec}, which is the first targeting DL-based \textbf{\underline{V}}olumetric \textbf{\underline{V}}ideo streaming \textbf{\underline{Rec}}onstruction attack scheme. We design latent diffusion models, the state-of-the-art point cloud generative approach, with a Gamma distribution, to construct and train a reconstruction attacker capable of reconstructing the original point clouds in high quality.
We also introduce a refinement algorithm, \textit{PCR}, to further improve the quality of reconstructed point clouds. 

To the best of our knowledge, VVRec is the first attempt to activate reconstruction attacks on DL-based volumetric video streaming. The contributions are as follows.
\begin{itemize}
    \item We introduce VVRec, a reconstruction attack scheme based on SOTA latent diffusion models with four NN modules, to construct and train a reconstruction attacker capable of reconstructing the original point clouds in high quality.
    \item We develop a \textit{PCR} algorithm to further improve the quality of the reconstructed point clouds.
    \item We compare VVRec on volumetric video datasets and attack seven victim models with other baselines. VVRec outperforms all in reconstruction accuracy.
\end{itemize}

\begin{figure}[t]
\centering
\includegraphics[width=0.46\textwidth]{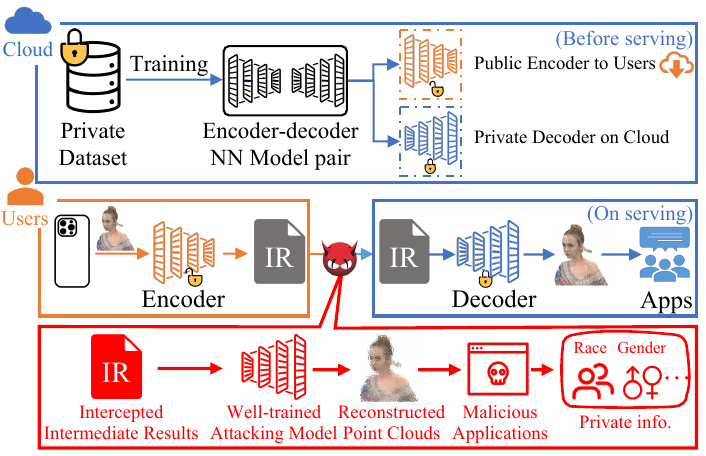}
\caption{Workflow of existing DL-based volumetric video streaming and privacy risks. }
\label{fig:intro}
\end{figure}

\section{Background and Related Work}
\subsection{Background}
\textbf{Volumetric Video Streaming.}
Existing volumetric video streaming methods can be categorized based on whether the transmission data constitutes the intermediate results of NN models or not~\cite{gao2022openpointcloud}:
(1) non-DL-based compression, e.g., G-PCC and V-PCC by MPEG~\cite{mpegpcc}, Pleno~\cite{jpeg}
and Draco~\cite{draco}.
(2) DL-based compression, e.g., PCGCv2~\cite{pcgcv2}, SparsePCGC~\cite{spcgc}, etc.

A typical DL-based volumetric video compression workflow is shown in Fig.~\ref{fig:intro}. As the 3D application service provider, the cloud constructs and trains a pair of NN models: an encoder and a decoder through a \textit{private} dataset. Only the well-trained encoder will be distributed to users, while the decoder is secured on the cloud. Upon capturing volumetric video, users compress it into intermediate results using this encoder. These results are then transmitted to the cloud and decoded by the secured decoder, ultimately decompressing to the original volumetric video. 

In this paper, we focus on recovering the volumetric video frame, i.e., point clouds, from the intercepted intermediate results during transmission in DL-based volumetric video compression.
Some researchers aim to enhance non-DL-based approaches for improved efficiency~\cite{akhtar2020point}, reduced distortion~\cite{zhang2023gpccpp}, and privacy protection~\cite{lu2023pagoda} using deep learning approaches. However, these methods are out of our scope because the transmitted data retains the same in non-DL-based approaches, which differs from the intermediate results of NN models in DL-based.

\noindent\textbf{Attacks on Volumetric Video Streaming.}
Attack on streaming is a serious threat that involves unauthorized access and interception of streaming data. Attackers can manipulate or impair transmission data, potentially jeopardizing confidentiality and integrity~\cite{ni2020security}.
Attackers may adopt various tactics, including faking as a colluded or corrupted server to extract users' private information~\cite{ren2023hcnct} or engaging in man-in-the-middle attacks~\cite{vladimirov2022security}. 
Among such attackers, reconstruction attackers target to recover the original input point cloud from the intermediate results. It attracts significant attention in deep learning~\cite{jegorova2022survey}. For instance, in language models, reconstruction attacks aim to reconstruct input text~\cite{brown2022does}, while in computer vision, they are to reconstruct input images~\cite{lu2022preva}. 

Reconstruction attacks usually target the application layer, which is quite different from attackers cracking encryption in the transport layer. For example, a malicious cloud manipulates users into sending their video data, even if the transmission is secured through HTTPS. While encryption methods can safeguard the data during transmission, they do not prevent the exposure of private information contained within the video content. To illustrate this kind of attack, Fig.~\ref{fig:intro} provides an example in red, demonstrating the attacking process of reconstructing point clouds from intercepted intermediate results. 
Before starting an attack, the attacker needs to train a corresponding attacking model capable of reconstructing the intercepted intermediate results. The attack commences by hijacking the intermediate results transmitted from the user to the cloud.
Reconstructed point clouds can be identified with sensitive information like facial features, gender, race, etc.

Various reconstruction attack approaches exist, including Supervised-based~\cite{dosovitskiy2016inverting}, GAN-based~\cite{spgan}, etc. However, the state-of-the-art generative solution, the latent diffusion model~\cite{rombach2022ldm}, has not been used in reconstruction attacks yet. 
Another similar attack is the attribute inference attack (AIA)~\cite{rigaki2023survey}, which extracts attributes of original inputs, e.g., gender, race, etc, from intermediate results. AIA only needs to recover a few points with attribute features.
Unlike AIA, reconstruction attacks demand more from attackers' capability as they need to recover the entire input data accurately.

\subsection{Related Work}
\textbf{Diffusion models (DMs)} have emerged as the most powerful generative models, showcasing remarkable performance in sample quality and good modality coverage~\cite{huang2021variational}, in area of image generation~\cite{anomaly,frido}, audio synthesis~\cite{kong2020diffwave}, etc, compared to other generative schemes, e.g., GANs, VAEs, etc.
DMs contain two phases: the \textit{forward} process to arbitrary noises, e.g., Gaussian noise~\cite{ho2020ddpm}, Poisson noise~\cite{xu2022poisson}, and the \textit{reverse} process to approximate the target distribution by denoising. 
Despite the success across diverse tasks, DMs suffer from sampling delays for thousands of network inference iterations, making them costly in real-world applications. However, it makes them particularly suitable for our attack scenario, where strict time constraints and computation capability are not decisive requirements. 
Generally, the Gaussian distribution in DM may lead to poor sample generation, e.g., the prior hole problem~\cite{sinha2021d2c,vahdat2021score}; however,
\textit{Denoising Diffusion Gamma Model} (DDGM)~\cite{nachmani2021ddgm} replaces it with the Gamma one, which has been proven to exhibit higher model capacity and improved convergence, particularly in high-dimensional data~\cite{chang2023design,croitoru2023diffusion}.

\noindent\textbf{Latent diffusion models (LDMs)} are initially introduced to enhance the efficiency of conventional DMs in high-resolution image generation~\cite{rombach2022ldm}. These models encode high-dimensional input data into a low-dimensional latent space and train DMs within this latent space. Those latents from the high-dimensional point cloud are named \textit{shape latent}~\cite{zeng2022lion}.
Samples generated in the latent space are then decoded back to their original high-dimensional data. 
In reconstruction attacks, the intermediate results happen to be in a low dimension, i.e., \textit{shape latents}, where the encoder of the victim model is the role of encoding high-dimensional to the low-dimensional latent space. All these features naturally contribute to the adoption of a new attacking scheme based on LDMs.

\section{Design}
\begin{figure*}[t]
    \centering
    \includegraphics[width=0.9\textwidth]{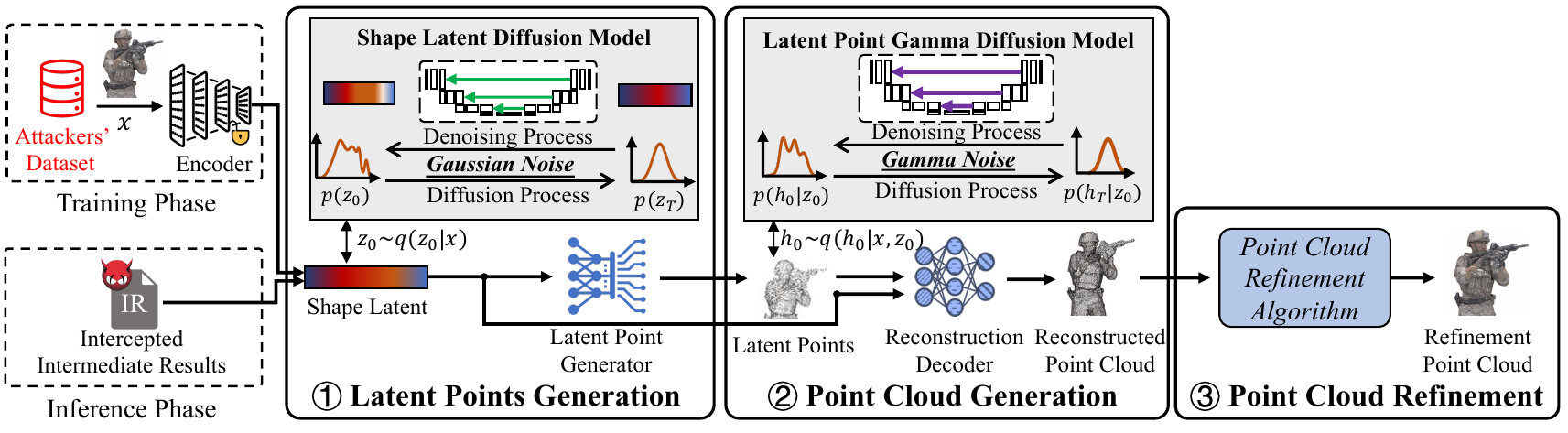}
    \caption{Workflow of VVRec activates a reconstruction attack, including training NN models in VVRec and reconstructing the given intercepted intermediate results to a point cloud.}
    \label{fig:design}
\end{figure*}
\subsection{Threat Model}
\textbf{Victim Model.} It is assumed that the cloud server owns a large volumetric video dataset or a specific domain dataset, which is inaccessible to others. The cloud will use it to train a DL-based compression model pair, including an encoder and a decoder. The decoder is private and will never be publicly exposed. The encoder will be distributed in public. Users capture point clouds, compress them by the encoder, and then transmit them to the cloud for decoding.

\noindent\textbf{Attack Model.} Following existing reconstruction attacks \cite{rigaki2023survey}, we assume the attacker owns sufficient computational resources and there is no latency constraint. Before the attack starts, the attacker accesses the encoder model in advance, possibly by spoofing attack. Although the attacker has volumetric data distinct from the cloud's, they can input their data into the encoder, generating intermediate results to train their attack model for recovering original inputs.
During the attack, the attacker intercepts intermediate results from victim users and utilizes the trained attacker model to reconstruct the original point cloud. The success of the attack is determined by the reconstructed point clouds achieving a high similarity to the original ones.

\subsection{Design Overview}
We present the overview of VVRec in Fig.~\ref{fig:design}. The objective of a well-trained VVRec is to reconstruct the point cloud from the intercepted intermediate results, i.e., \textit{shape latent}, in high quality. The source of the shape latent varies during training and inference. During training, the shape latent is derived from a dataset created by the attacker inputting their volumetric data into the compression encoder provided by the target victim model. In the attack, i.e., during the inference phase, the intercepted intermediate result is the shape latent that has been encoded by the victim user. In general, there are three major steps in VVRec:

\noindent~~ $\bullet$ \textbf{Latent Points Generation} (\S~\ref{sec:design-lpg}): The shape latent is initially processed through the \textit{Shape Latent Diffusion Model (SLDM)}, where it undergoes a forward diffusion process followed by a reverse denoising process to sample a new shape latent. Subsequently, the updated shape latent is transmitted to the \textit{Latent Point Generator (LPG)} for encoding into a point cloud format, named \textit{latent points}.

\noindent~~ $\bullet$  \textbf{Point Cloud Generation} (\S~\ref{sec:design-pcg}): The latent points from LPG will first be sent to the \textit{Latent Point Gamma Diffusion Model (LPGDM)} and that undergoes a forward diffusion process using a \textit{Gamma} distribution and a reverse denoising process to sample a new latent point cloud. The resulting new latent point cloud, along with the shape latent, serves as the input to the \textit{Reconstruction Decoder (RD)}, producing the \textit{reconstructed point cloud}.

\noindent~~ $\bullet$  \textbf{Point Cloud Refinement} (\S~\ref{sec:design-ref}): To enhance the quality of the reconstructed point cloud generated by RD, a \textit{Point Cloud Refinement (PRC) Algorithm} is designed to convert the point cloud into a mesh format, and resample to reconstruct a high-quality refinement point cloud.

\subsection{Latent Points Generation} \label{sec:design-lpg}
We first model the volumetric video frame, i.e., point cloud, defined as $\pointcloud \in \mathbb{R}^{(3+c)\times N}$, where $N$ is the number of points with coordinates $xyz \in \mathbb{R}^3$. Here, $c$ is the color representation, such as points are in RGB format if $c=3$. $c=0$ indicates that the volumetric video lacks color (or transmits color in the other way). The shape latent, denoted as $\shapelat_0 \in \mathbb{R}^{D_\shapelat}$ where $D_\shapelat$ is the dimension of $\shapelat_0$ determined by the victim model setup. Latent point is in point cloud format, denoted as $\latpoint_0 \in \mathbb{R}^{(3+c+D_\latpoint)\times N}$, with $N$ points with geometry coordinates $xyz \in \mathbb{R}^3$, and $D_\latpoint$ is the additional latent feature dimension. 

\noindent~~$\bullet$~~\textit{\textbf{Shape Latent Diffusion Model (SLDM).}} Shape latent $\shapelat_0$ will be first sampled from the SLDM. Specifically, $\shapelat_0$ follows a forward process to add Gaussian distribution noise in DMs.
\begin{equation}~\label{eq:dm_fw}
\shapelat_t = \sqrt{\bar{\alpha}_t}\shapelat_0+\sqrt{1-\bar{\alpha}_t}\boldsymbol{\epsilon},\ t=1,2,...,T, \boldsymbol{\epsilon}\sim \mathcal{N}(\textbf{\textit{0}},\textbf{\textit{I}})
\end{equation}
where $\alpha_t=1-\beta_t$, $\bar{\alpha}_t=\prod^t_{i=1}\alpha_i$.
$T$ denotes the number of diffusion steps and is set to $T_{SL}$ in SLDM. The noise scheduling sequence $\{\beta_t\}$ is chosen such that the chain approximately converges to a standard Gaussian distribution~\cite{ho2020ddpm}, i.e., $\mathcal{N}(\textbf{\textit{0}},\textbf{\textit{I}})$, after $T$ steps, if $T$ is given large enough. The reverse process of SLDM is then defined as:
\begin{equation}~\label{eq:dm_bw}
    \shapelat_{t-1} = \frac{1}{\sqrt{\alpha_t}}(\shapelat_t-\frac{\beta_t}{\sqrt{1-\bar{\alpha}_t}}\boldsymbol{\epsilon_\theta}(\shapelat_t,t)), t=1,2,...,T-1
\end{equation}
SLDM is designed to learn the reverse process with parameters $\boldsymbol{\theta}$ that inverts the forward diffusion by predicting $\boldsymbol{\epsilon_\theta}(\shapelat_t,t)$. The new shape latent $\shapelat_0$ is sampled step by step from $\{\shapelat_T, \shapelat_{T-1},...\}$ by iteratively calculating Eq.~\ref{eq:dm_bw}. 
The simplified loss function for SLDM is defined as follows:
\begin{equation}~\label{dm:loss}    L_\shapelat(\boldsymbol{\theta})=\mathbb{E}_{\shapelat_t|\pointcloud,\boldsymbol{\epsilon}\sim \mathcal{N}(\boldsymbol{0},\boldsymbol{I})}||\boldsymbol{\epsilon}-\boldsymbol{\epsilon_\theta}(\shapelat_t,t)||^2_2
\end{equation}
To support such functions, we follow \cite{zhou} and build SLDM based on PVCNN \cite{pvcnn}. To model the distribution of shape latents, we add ResNet-like components into the SLDM with residual connections.

\noindent~~$\bullet$~~\textit{\textbf{Latent Point Generator (LPG).}} 
To construct latent points $\latpoint_0$ in point cloud format from the sampled shape latent $\shapelat_0$, we design LPG based on PVCNN model to support such functions with parameters $\boldsymbol{\phi}$ as well. LPG takes $\latpoint_0$ as input. To enable the conditioning on shape latents, we add adaptive Group Normalization modules to the original PVCNN model.
Finally, $\latpoint_0|\shapelat_0$ is obtained with the dimension of $3+c+D_\latpoint$. 
As LPG is intended to be trained in conjunction with the Reconstruction Decoder in the next section, we will defer the discussion of its loss function later.

\subsection{Point Cloud Generation}~\label{sec:design-pcg}
In this section, the latent points $\latpoint_0$, generated by LPG are initially processed by the LPGDM to sample new latent points. The resulting new latent points with the shape latent as conditions, serve as the input to the RD, producing the reconstructed point cloud $\pointcloud'$.

\noindent~~$\bullet$~~\textit{\textbf{Latent Point Gamma Diffusion Model (LPGDM).}} 
Similar to Eq.~\ref{eq:dm_fw} SLDM, the sampling process of $\latpoint_0$ starts with a forward process but incorporates Gamma distribution noise~\cite{nachmani2021ddgm} instead of Gaussian, as follows:
\begin{equation}~\label{eq:gm_fw}
\latpoint_t =\sqrt{\bar{\alpha_t}}\latpoint_{0}+(\bar{g}_t-\bar{k}_t\theta_t), t=1,2,...,T-1
\end{equation}
where $\bar{g}_t\sim\Gamma(\bar{k}_t,\theta_t)$, $\theta_t=\sqrt{\bar{\alpha_t}\theta_0}$, 
$\bar{k}_t=\sum^{t}_{i=1}\frac{\beta_t}{\alpha_t\theta^2_0}$. 
$\theta_0$ is the initial Gamma distribution scale hyperparameter.
The reverse process is then defined as follows :
\begin{equation}~\label{eq:gm_bw}\footnotesize
\begin{split}
    \latpoint_{t-1} &= \frac{1}{\sqrt{\alpha_t}}(\latpoint_t-\frac{\beta_t}{\sqrt{1-\bar{\alpha}_t}}\boldsymbol{\epsilon_\psi}(\latpoint_t,\shapelat_0,t)) \\
    &+\beta_t\frac{\gamma_{t-1}-\theta_{t-1}\bar{k}_{t-1}}{\sqrt{1-\bar{\alpha}_t}},~t=1,2,...,T-1
\end{split}
\end{equation}
where $\gamma_t\sim\Gamma(\theta_t.\bar{k}_t)$, and $T:=T_{LPG}$ is the diffusion steps number.

LPGDM is designed to learn the reverse process with parameters $\boldsymbol{\psi}$ by predicting $\boldsymbol{\epsilon_\psi}(\latpoint_t|\shapelat_0,t)$. LPGDM applies PVCNN structure to efficiently model the reverse process.

The latent point clouds $\latpoint_0|\shapelat_0$ are iteratively sampled by stepwise calculations according to Eq.~\ref{eq:gm_bw}. The loss function of LPGDM is defined as follows:
\begin{equation}~\label{gm:loss}
L_\latpoint(\boldsymbol{\psi})=\mathbb{E}_{\latpoint_t|\shapelat_0,\bar{g}_t\sim\Gamma(\bar{k}_t,\theta_t)}||\frac{(\bar{g}_t-\bar{k}_t\theta_t)}{\sqrt{1-\bar{\alpha}_t}}-\boldsymbol{\epsilon_\psi}(\latpoint_t,t)||^2_2
\end{equation}

\noindent~~$\bullet$~~\textit{\textbf{Reconstruction Decoder (RD).}} 
Following the sampling of latent points $\latpoint_0|\shapelat_0$ by LPGDM, a reconstruction decoder model with parameters $\boldsymbol{\xi}$ is devised to reconstruct the point cloud $\pointcloud'$ using the input latent points and shape latent $\shapelat_0$. 
The design of RD is constructed as a VAE based on PVCNN. Different from the LPG, the output of RD conditions on two inputs, $\shapelat_0$ and $\latpoint_0$. To accommodate the conditioning on two latent encodings, we replace the original group normalization in PVCNN with adaptive group normalization (AdaGN). To be specific, the shape latent $\shapelat_0$ is cut into factor and bias, and the latent point $\latpoint_0$ is group-normalized. Finally, the product of group-normalized $\latpoint_0$ and factor plus the bias is the output of the AdaGN layer. The reconstructed point cloud $\pointcloud'$ is the weighted summation of RD output and latent points.
The training of RD is conducted in conjunction with LPG. The loss function is designed based on the maximization of a modified variational lower bound on the data log-likelihood (ELBO)~\cite{hoffman2016elbo}:
\begin{equation}~\label{eq:rd}\footnotesize
\begin{split}
    &L_{\text{ELBO}}(\boldsymbol{\phi},\boldsymbol{\xi})=\mathbb{E}_{\pointcloud',\shapelat_0,\latpoint_0|\shapelat_0} [ \text{log} p_{\boldsymbol{\xi}}(\pointcloud'|\latpoint_0,\shapelat_0)- \\ 
    &\lambda_\shapelat D_{\text{KL}}(q_{\boldsymbol{\phi}}(\shapelat_0)|p(\shapelat_0))-\lambda_\latpoint D_{\text{KL}}(q_{\boldsymbol{\phi}}(\latpoint_0|\shapelat_0)|p(\latpoint_0))]. 
\end{split}
\end{equation}
where $p(\latpoint_0), p(\shapelat_0)\sim\mathcal{N}(\mathbf{0},\mathbf{I})$, $p_{\boldsymbol{\xi}}(\pointcloud'|\latpoint_0,\shapelat_0)$ denotes RD. Here, $\lambda_\shapelat$ and $\lambda_\latpoint$ are hyperparameters balance reconstruction accuracy and K-L regularization $D_{\text{KL}}$~\cite{zeng2022lion}.

\begin{algorithm}[t]
\KwIn{Reconstructed Point Cloud $\pointcloud'\in\mathbb{R}^{3+c+N}$, Maximum \#points $N_{max}$, Maximal quality gradient $\delta_{max}$}
\KwOut{Refinement Point Cloud $\pointcloud''\in\mathbb{R}^{3+c+N'}$}
$M \leftarrow SAP(\pointcloud')$ \;
\While{$\#\mathbf{y} \leq N_{max}$}{
$\mathbf{y} \leftarrow PDS(M, d_0)$ \;
\If{$ p2plane\_PSNR(\mathbf{y},\mathbf{y_{last}})\leq\delta_{max}$}{
Increase upsampling precision $d_0$ \;
$\mathbf{y_{last}}\leftarrow\mathbf{y}$ \;
}
\Else{
\textbf{Break} \;
}
}
Output $\pointcloud''$.
\caption{Point Cloud Refinement Algorithm}
\label{alg:pcr}
\end{algorithm}

\subsection{Point Cloud Refinement}~\label{sec:design-ref}
To generate a point cloud of high quality, we develop a Point Cloud Refinement Algorithm (PCR) as shown in Algorithm~\ref{alg:pcr} to refine the reconstructed point cloud. 
It is able to smooth the point cloud surface and simultaneously cancel out the noisy points from the LPGDM.
In Line 1, We initially utilize a well-trained \textit{Shape As Points (SAP)}~\cite{peng2021sap}, which efficiently extracts high-quality watertight meshes from point clouds based on Differentiable Poisson Surface Reconstruction algorithm. It establishes relationships among independent points in $\pointcloud'$ to mesh, denoted $M$.
From Lines 2 to 8, we iteratively generate point clouds by increasing sampling precision $d_0$ via Poisson Disk Sampling (PDS)~\cite{bowers2010parallel} and calculating the similarity (p2plane-PSNR) between the sampled point clouds and the last generated one. We will yield a refinement point cloud, denoted as $\pointcloud''$ with $N'$ point quantity until the similarity is higher than the maximal quality gradient $\delta_{max}$ or exceeding the maximal point number $N_{max}$. 

\section{Implementation and Evaluation}\label{sec:eva}

\begin{figure}[t]
\centering
\includegraphics[width=0.455\textwidth]{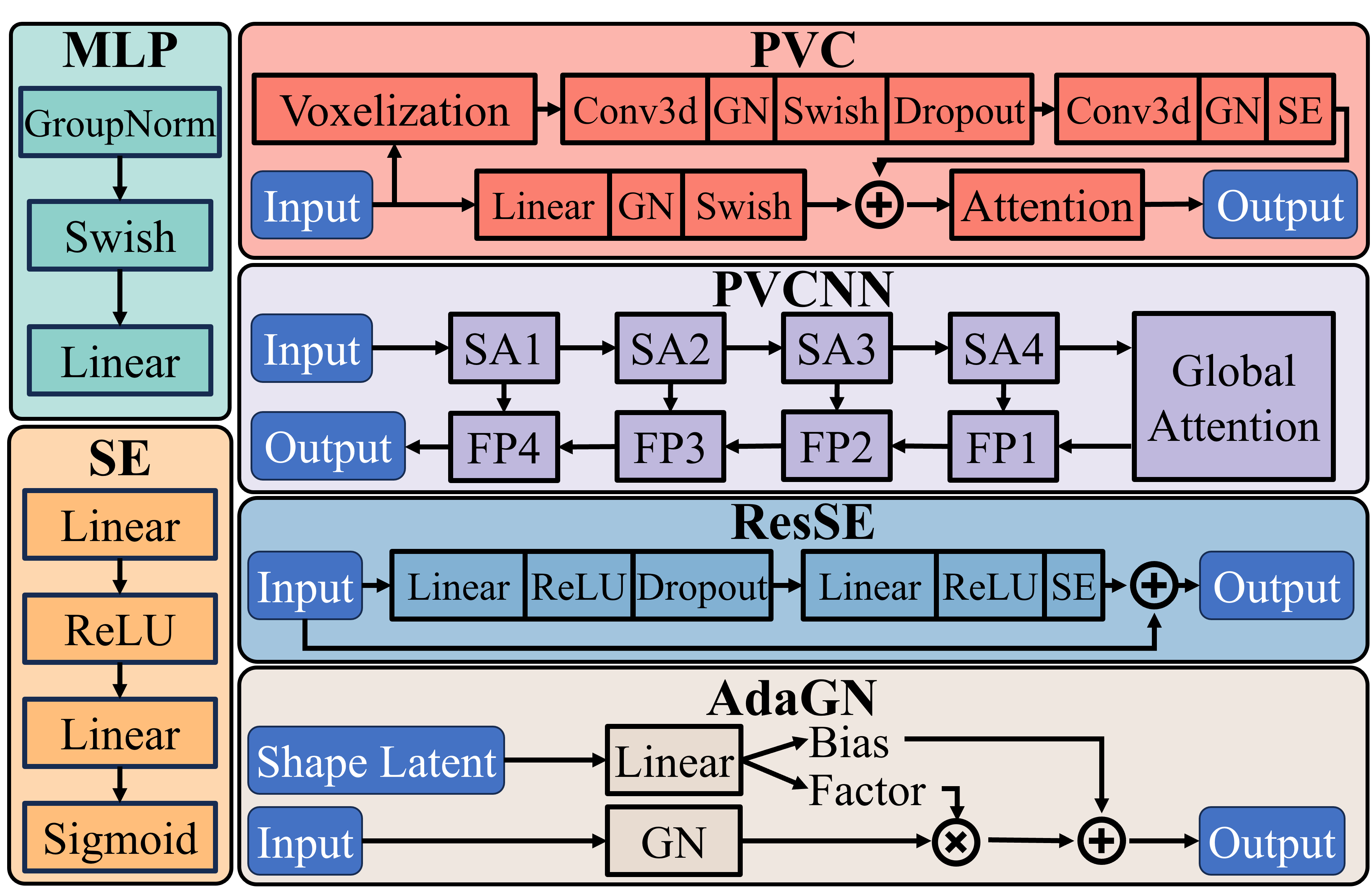}
\caption{Basic blocks of VVRec.}
\label{fig_app}
\end{figure}

\subsection{Implementation}\label{sec:eva_imp}
We establish VVRec by building the two diffusion models, SLDM and LPDGM, plus an LPG and an RD on PVCNN \cite{pvcnn} for the point cloud. Point cloud refinement is achieved through the Shape-as-Points (SAP) \cite{peng2021sap}, which generates a mesh from the reconstructed point cloud. Utilizing Alg.~\ref{alg:pcr}, we produce a refined point cloud. The training of VVRec and the two diffusion models are developed based on HuggingFace. Victim models are supported by OpenPointCloud project~\cite{gao2022openpointcloud}. 

\noindent\textbf{Training details.} There are two stages of training VVRec. Firstly, we train the LPG with RD by optimizing the loss function in Eq.~\ref{eq:rd}. During this process, the latent point $\latpoint_0$ is generated from the LPG model. The reconstructed point cloud $\pointcloud'$ is sampled from its posterior distribution conditioning on the shape latent $\shapelat_0$ and the latent point $\latpoint_0$ estimated by the RD. 
Secondly, we freeze the LPG and RD models to train SLDM and LPGDM on the shape latent $\shapelat_0$ and latent point $\latpoint_0$, minimizing the loss functions defined in Eq.~\ref{dm:loss} and \ref{gm:loss} respectively. The number of steps, $T_{SL},\ T_{LPG}$ in SLDM and LPGDM are set to 1000. We utilize Adam optimizer with the learning rate of LPG/RD, SLDM, and LPGDM as 10e-3, 10e-4, and 10e-4, respectively. The value of $N$ is configured as 2e4. The number of parameters is 174M in total, and the computation complexity is 1.05 TFLOPS. The overall training time in a 32 batch size is 500 hours.

\noindent\textbf{Details of model architecture.} The blocks used for implementing modules of VVRec are displayed in Fig.~\ref{fig_app}. Here, multi-layer perceptron (MLP), point-voxel convolution (PVC), set abstraction (SA), and feature propagation (FP) are the four building blocks of our PVCNN-like modules. For the details of SA and FP modules, please refer to PointNet++\cite{pointnet++}. The squeeze-and-excitation (SE) module is inherited from \cite{squeeze}, and the ResSE module denotes the ResNet-like residual module with the SE layer. The adaptive group normalization (AdaGAN) layer is designed for conditioning the shape latent. The details of VAE models and diffusion models are displayed as follows:

$\bullet$  \textbf{VAE Models} Our latent point generator (LPG) and reconstruction decoder (RD) are constructed as VAEs. Both LPG and RD consist of PVC layers, the global attention layer, MLP layers, the dropout layer, and the linear output layer. All dropout layers in the VAEs are implemented with a dropout probability of 0.1. To fulfill conditioning on the shape latents, we replace them with the AdaGN layers. The initial weight scale is 0.1. We also add residual connections in various layers. In the LPG, the raw input point cloud plus the product of 0.01 and the mean of the predicted latent point form the final output. In the RD, the final output is the summation of the sampled latent point cloud and 0.01 times the predicted point cloud.

$\bullet$ \textbf{Diffusion Models} 
We illustrate the details of the Shape Latent Diffusion Model (SLDM) and Latent Point Gamma Diffusion Model (LPGDM) in Tab.~\ref{tab_appendix3} and Tab.~\ref{tab_appendix4}, respectively. Similar to the VAE models, a dropout probability of 0.1 is adopted for all dropout layers in the SLDM and LPGDM, and we replace all GN layers with AdaGN layers to realize conditioning on the shape latent. For the AdaGN layers, the initial weight scale is 0.1. For all SA and FP layers, the point features are concatenated with time embeddings to achieve the diffusion-denoising process.

\begin{table}[t]
\centering
\small
\setlength{\tabcolsep}{1.75pt}{
\begin{tabular}{c}
\bottomrule[1.25pt]
Input: latent points $h_t$ at \textit{t}, shape latent $z_0$ of dimension ($1\times D_z$) \\ \hline
Output: new shape latent $z_0$ of dimension ($1\times D_z$)\\ \hline
\hline
Time embedding layer: \\
Sinusoidal embedding dimension = 128 \\
Linear (128, 512) \\
LeakyReLU (0.1) \\
Linear (2048) \\ \hline
\hline
Linear (128, 2048) \\
Addition (linear output, time embedding) \\
ResSE (2048, 2048) x 8 \\
Linear (2048, 128) \\
\bottomrule[1.25pt]
\end{tabular}
\caption{Architecture of SLDM.}
\label{tab_appendix3}
}
\end{table}

\begin{table}[t]
\centering
\setlength{\tabcolsep}{1.55pt}{
\small
\begin{tabular}{ccccc}
\bottomrule[1.25pt]
\multicolumn{5}{c}{Input: latent points $h_t$ at \textit{t}, shape latent $z_0$} \\ \hline
\multicolumn{5}{c}{Output: new latent points $h_0$} \\ \hline
\hline
\multicolumn{5}{c}{Time embedding layer} \\ \hline
\multicolumn{5}{c}{Sinusoidal embedding dimension = 64} \\ 
\multicolumn{5}{c}{Linear (64, 64)} \\ 
\multicolumn{5}{c}{LeakyReLU (0.1)} \\ 
\multicolumn{5}{c}{Linear (64, 64)} \\ \hline
\hline
 & SA1 & SA2 & SA3 & SA4 \\ \hline
 \# PVC Layers & 2 & 1 & 1 & N/A \\
 \# PVC hidden dimension & 32 & 64 & 128 & N/A \\
 \# PVC voxel grid size & 32 &16 &8 &N/A \\
\# Grouper center &1024 &256 &64 &16 \\
\# Grouper radius &0.1 &0.2 &0.4 &0.8 \\
\# Grouper neighbors &32 &32 &32 &32 \\
\# MLP layers &2 &2 &2 &3 \\
\# MLP output dimension &32,32 &64,128 &128,128 &128,128,128 \\
\# Attention dimension &N/A &128 &N/A &N/A \\ \hline
\hline
\multicolumn{5}{c}{Global attention, hidden dimension of 256} \\ \hline
\hline
 & FP1 & FP2 & FP3 & FP4 \\ \hline
\# MLP layers &2 &2 &2 &3 \\
\# MLP output dimension &128,128 &128,128 &128,128 &128,128,64 \\
\# PVC layers &3 &3 &2 &2 \\
\# PVC hidden dimension &128 &128 &128 &64 \\
\# PVC voxel grid size &8 &8 &16 &32 \\ \hline
\hline
\multicolumn{5}{c}{MLP: (64,128)} \\
\multicolumn{5}{c}{Dropout} \\
\multicolumn{5}{c}{Linear: (128, $3+c+D_h$)} \\
\bottomrule[1.25pt]
\end{tabular}}
\caption{Architecture of LPGDM.}\label{tab_appendix4}
\end{table}

\begin{table*}[t]
\centering
\setlength{\tabcolsep}{2pt}{
\small
\begin{tabular}{c|c|l|cc|cc|cc|cc|cc}
\bottomrule[1.25pt]
\multirow{2}{*}{\textbf{Attacker Dataset}} & \multirow{2}{*}{\textbf{Victim Dataset}} & \textbf{Victim Encoder} & \multicolumn{2}{c|}{\textbf{\textit{PCGCv1}}} & \multicolumn{2}{c|}{\textbf{\textit{PCGCv2}}} & \multicolumn{2}{c|}{\textbf{\textit{SparsePCGC}}} & \multicolumn{2}{c|}{\textbf{\textit{Pcc-geo-cnn-v1}}} & \multicolumn{2}{c}
{\textbf{\textit{Pcc-geo-cnn-v2}}} \\ \cline{3-13}
& & \textbf{Metrics} & M1 & M2 & M1 & M2& M1 & M2& M1 & M2 & M1 & M2 \\ \bottomrule[1.25pt]
\multirow{6}{*}{MVUB} & \multirow{6}{*}{8iv2} & \textcircled{1} \textit{Optimal} & 72.02  & 76.85 & 73.75 & 77.61 & 76.88 & 80.97 & 68.16 & 72.07 &70.78 & 74.39  \\  \cline{3-13}
& & \textcircled{2} \textit{Supervised} & 45.62 & 50.01 & 41.03  & 45.75  & 40.26 & 45.06 & 46.81 & 51.36 & 41.57 & 46.01 \\ 
& & \textcircled{3} \textit{SP-GAN} & 50.15 & 54.95 & 45.71 & 50.04 & 44.36 & 48.91 & 48.15 & 52.93 & 45.57 & 50.81 \\
& &  \textcircled{4} \textit{VG-VAE}  & 48.05 & 52.78 & 43.61 & 47.84 & 42.16 & 46.51 & 46.02 & 50.73 & 43.67 & 48.95 \\
& &  \textcircled{5} \textit{PD-Flow}  & 47.25 & 51.48 & 41.71 & 46.92 & 41.33 & 45.92 & 44.68 & 49.03 & 42.66 & 47.80 \\
& & \textcircled{6} \textit{VVRec} & \textbf{58.75} & \textbf{63.52} & \textbf{56.21} & \textbf{60.82} & \textbf{53.72} & \textbf{58.01} & \textbf{59.24} & \textbf{63.78} & \textbf{57.01} & \textbf{62.08} \\ \bottomrule[1.25pt]
\multirow{6}{*}{MVUB} & \multirow{6}{*}{Owlii} & \textit{Optimal} & 72.18  & 76.45 & 73.48 & 77.56 & 76.78 & 80.57 & 67.85 & 72.12 &70.45 & 74.03  \\  \cline{3-13}
& & \textit{Supervised} & 45.68 & 50.02 & 41.07  & 45.73  & 40.29 & 45.11 & 46.88 & 51.29 & 41.58 & 46.07 \\ 
& & \textit{SP-GAN} & 50.06 & 54.98 & 45.70 & 50.09 & 44.28 & 48.98 & 48.13 & 52.92 & 45.61 & 50.84 \\
& & \textit{VG-VAE}  & 48.09 & 52.77 & 43.59 & 47.86 & 42.17 & 46.53 & 46.06 & 50.74 & 43.66 & 48.92 \\
& &\textit{PD-Flow}  & 47.24 & 51.45 & 41.73 & 46.96 & 41.23 & 46.01 & 44.72 & 49.00 & 42.69 & 47.77 \\
& & \textit{VVRec} & \textbf{58.72} & \textbf{63.49} & \textbf{56.22} & \textbf{60.80} & \textbf{53.75} & \textbf{58.04} & \textbf{59.27} & \textbf{63.75} & \textbf{57.02} & \textbf{62.15} \\ \bottomrule[1.25pt]
\multirow{6}{*}{MVUB} & \multirow{6}{*}{UVG-VPC} & \textit{Optimal} & 73.08  & 77.45 & 74.56 & 78.21 & 77.36 & 81.73 & 69.12 & 73.01 &71.45 & 74.98  \\  \cline{3-13}
& & \textit{Supervised} & 45.01 & 49.12 & 40.48  & 45.06  & 39.72 & 44.29 & 45.93 & 50.88 & 41.02 & 45.62 \\ 
& & \textit{SP-GAN} & 49.45 & 54.16 & 44.99 & 49.78 & 43.89 & 48.20 & 47.45 & 52.33 & 45.01 & 50.14 \\
& & \textit{VG-VAE}  & 47.56 & 52.08 & 43.02 & 47.17 & 41.65 & 45.97 & 45.48 & 50.09 & 43.04 & 48.17 \\
& &\textit{PD-Flow}  & 46.91 & 50.89 & 41.04 & 46.13 & 40.56 & 45.14 & 43.98 & 48.73 & 42.05 & 47.17 \\
& & \textit{VVRec} & \textbf{58.05} & \textbf{62.89} & \textbf{55.54} & \textbf{60.08} & \textbf{53.16} & \textbf{57.48} & \textbf{58.78} & \textbf{63.26} & \textbf{56.25} & \textbf{61.67} \\ \bottomrule[1.25pt]
\multirow{6}{*}{Owlii} & \multirow{6}{*}{8iv2} & \textit{Optimal} & 72.02  & 76.85 & 73.75 & 77.61 & 76.88 & 80.97 & 68.16 & 72.07 &70.78 & 74.39  \\  \cline{3-13}
& & \textit{Supervised} & 50.03 & 54.49 & 45.79 & 50.46 & 44.31 & 49.20 & 49.11 & 53.64 & 46.66 & 51.37 \\
& & \textit{SP-GAN} & 52.18 & 56.36 & 48.86 & 53.72 & 47.77 & 52.57 & 51.21 & 55.67 & 48.77 & 53.67 \\
& & \textit{VG-VAE} & 50.16 & 54.38 & 46.88 & 51.71 & 45.76 & 50.47 & 49.03 & 53.36 & 46.97 & 51.87  \\
& & \textit{PD-FLow} & 49.19 & 53.06 & 44.86 & 50.63 & 44.57 & 49.48 & 48.20 & 52.47 & 44.97 & 49.86 \\
& & \textit{VVRec} & \textbf{60.28} & \textbf{64.70} & \textbf{58.01} & \textbf{62.43} & \textbf{55.88} & \textbf{60.97} & \textbf{59.97} & \textbf{64.70} & \textbf{58.81} & \textbf{63.27} \\ \bottomrule[1.25pt]
 \multirow{6}{*}{UVG-VPC} & \multirow{6}{*}{8iv2} & \textit{Optimal} & 72.02  & 76.85 & 73.75 & 77.61 & 76.88 & 80.97 & 68.16 & 72.07 &70.78 & 74.39  \\  \cline{3-13}
& & \textit{Supervised} & 49.93 & 54.27 & 45.82 & 50.63 & 44.01 & 49.14 & 48.73 & 53.03 & 46.45 & 51.27 \\
& & \textit{SP-GAN} & 52.33 & 56.04 & 47.99 & 53.02 & 48.07 & 52.65 & 51.01 & 55.68 & 48.72 & 53.69 \\
& & \textit{VG-VAE} & 50.03 & 54.16 & 46.67 & 51.73 & 44.98 & 50.26 & 51.31 & 55.89 & 46.98 & 51.89 \\
& & \textit{PD-FLow} & 49.02 & 53.14 & 44.97 & 50.69 & 44.29 & 49.37 & 48.01 & 52.38 & 45.08 & 49.98 \\
& & \textit{VVRec} & \textbf{60.33} & \textbf{64.85} & \textbf{57.89} & \textbf{62.17} & \textbf{55.69} & \textbf{60.87} & \textbf{60.32} & \textbf{65.27} & \textbf{58.64} & \textbf{63.43} \\ \bottomrule[1.25pt]
\end{tabular}
\caption{Reconstruction quality results measured by p2p-PSNR (M1$\uparrow$) and p2plane-PSNR (M2$\uparrow$) with bitrate 0.4bpp. }\label{tab_overall}
}
\end{table*}

\subsection{Experimental Setup}
\noindent\textbf{Testbed.} Our experiments adopt a powerful workstation featuring dual NVidia RTX 4090 GPUs with a total graphic memory of 48GB, along with an Intel i9 CPU. 

\noindent\textbf{Datasets.} Four datasets are adopted, which are: 1) \textit{8i Voxelized Full Bodies (8iv2)}~\cite{8iv2}, 2) \textit{Owlii Dynamic Human Textured (Owlii)}~\cite{Owlii}, 3) \textit{Microsoft Voxelized Upper Bodies (MVUB)}~\cite{MVUB}, 4) \textit{UVG-VPC}~\cite{uvgvpc}. All of them are certified by the JPEG as the benchmark dataset for volumetric video streaming.

\noindent\textbf{Victim models.} We use five DL-based volumetric video streaming models as our attacking targets, including 
1) PCGCv1 \cite{pcgcv1},
2) PCGCv2 \cite{pcgcv2},
3) SparsePCGC \cite{spcgc},
4) Pcc-geo-cnn-v1 \cite{quach2019learning},
5) Pcc-geo-cnn-v2 \cite{pccv2}, which are lossy models.
6) VoxelDNN \cite{voxeldnn} and 
7) MSVoxelDNN \cite{msvoxeldnn} are lossless models.
Without further specification, the bit rate of these victim models is set to 0.4 bpp. The evaluation results of VoxelDNN and MSVoxelDNN, and other bit rates are listed in the Appendix.

\noindent\textbf{Baselines.} We compare VVRec performance with four reconstruction attack methods and the decoder coupled with the victim encoder:
$\bullet$ \textbf{\textit{Supervised }} has symmetric NN structure to the victim encoder model. The input point clouds of the victim encoder are the training labels and the output intermediate results are the training data.
$\bullet$ \textbf{\textit{SP-GAN~\cite{spgan}}} trained a generator and a discriminator where the generator input intermediate results from the victim encoder and output reconstructed point cloud and adversarial training with the discriminator.
$\bullet$ \textbf{\textit{VG-VAE~\cite{vgvae}}} applies a hierarchical VAE. 
$\bullet$ \textbf{\textit{PD-Flow~\cite{pdflow}}} is a point cloud denoising framework with normalizing flows.
$\bullet$ \textbf{\textit{Optimal }}is the decoder coupled with the victim encoder and serves as the upper bound of reconstruction performance.

\noindent\textbf{Evaluation metrics.} We use the point-to-point Peak Signal-to-Noise Ratio (p2p-PSNR) and point-to-plane PSNR (p2plane-PSNR) \cite{tian2017geometric}, denoted as M1 and M2, respectively. The higher value represents higher similarity and better reconstruction performance.

\begin{figure}[t]
\centering
\includegraphics[width=0.425\textwidth]{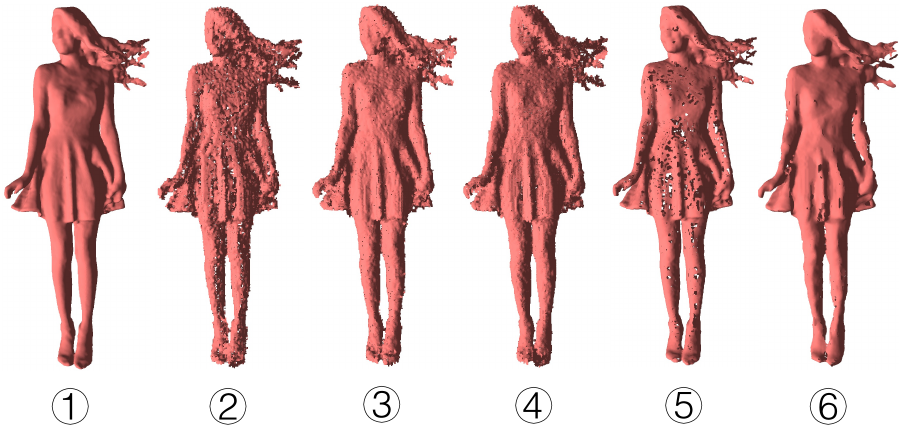}
    \caption{Visualization of reconstruction result of VVRec and baselines on trace \textit{redandblack@8i} during attacking victim encoder PCGCv1 on above Table 1.}
    \label{fig_overall}
\end{figure}

\begin{figure*}[!th]
\centering
\begin{minipage}[t]{\linewidth}
\centering
\subfigure[Original]{
\includegraphics[width=0.151\textwidth]{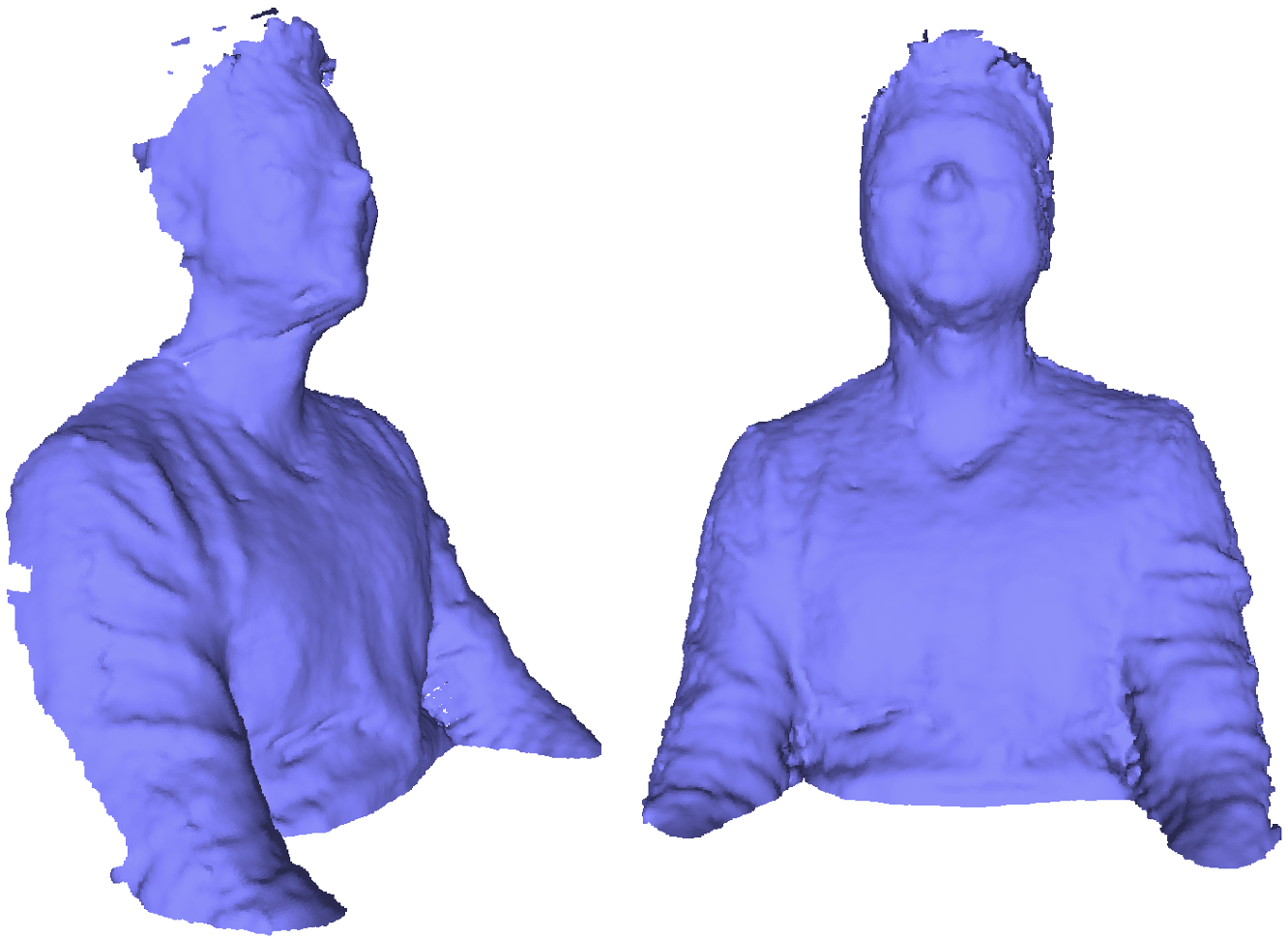}}
\subfigure[Paired decoder]{
\includegraphics[width=0.151\textwidth]{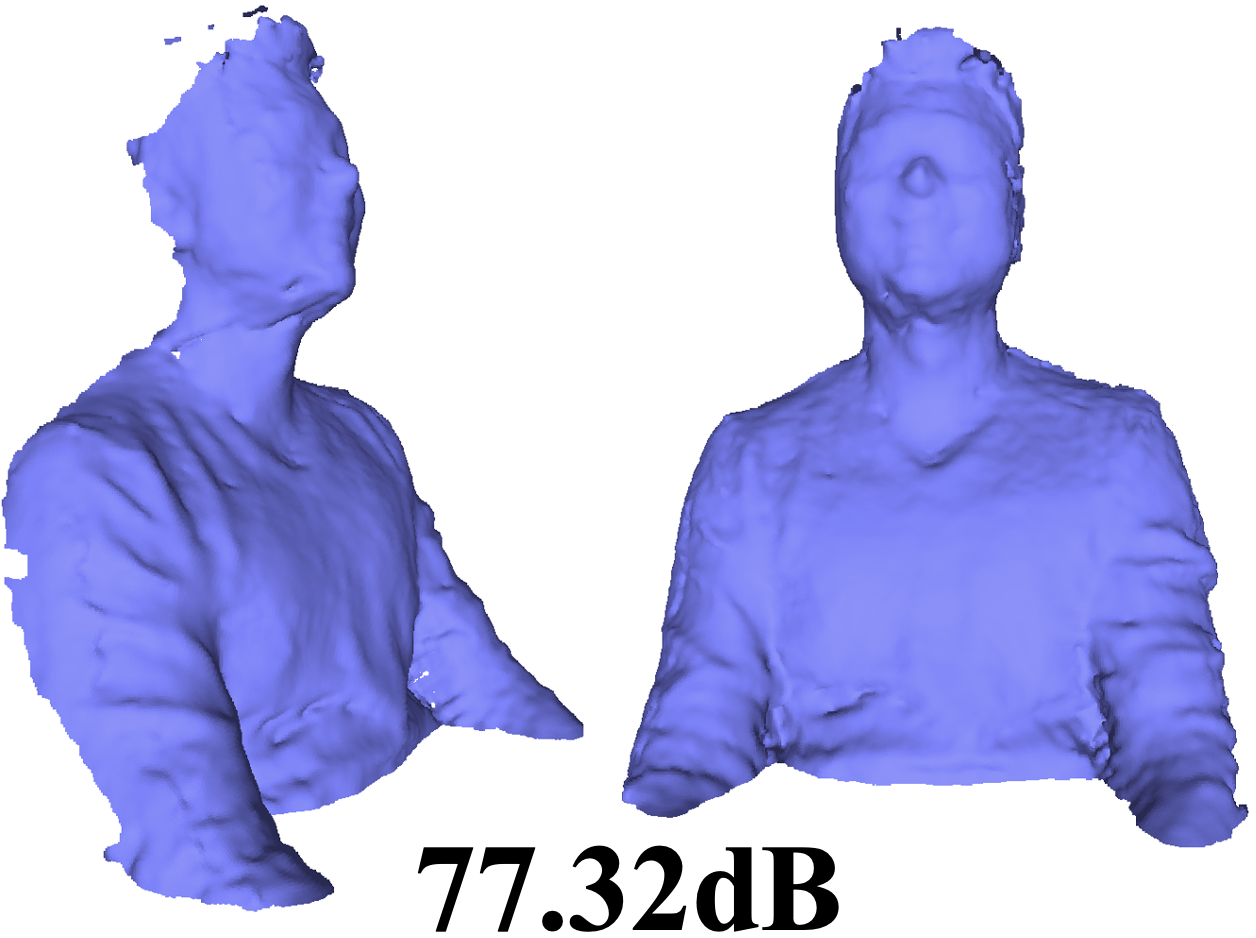}}
\subfigure[VVRec-A]{
\includegraphics[width=0.151\textwidth]{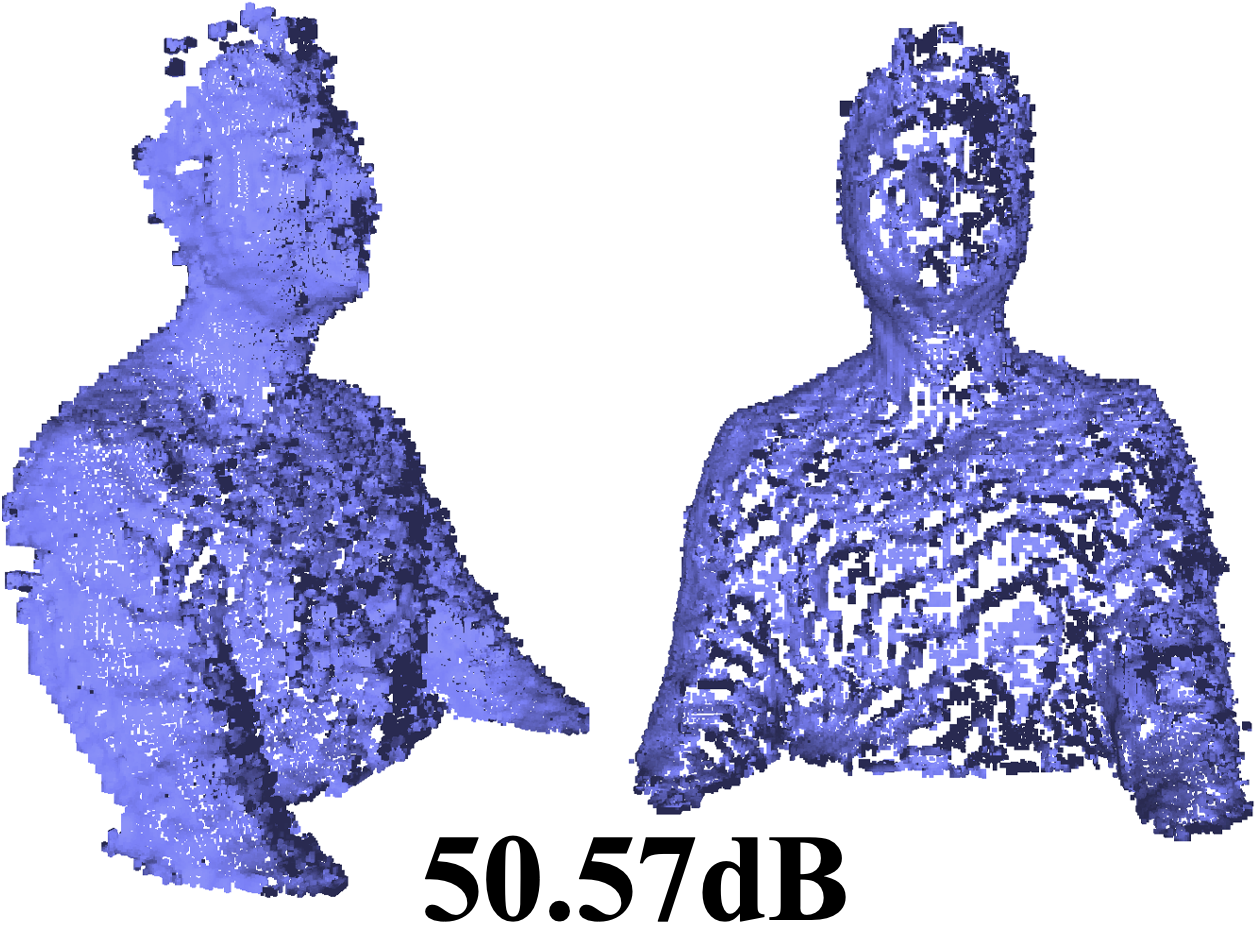}}
\subfigure[VVRec-B]{
\includegraphics[width=0.151\textwidth]{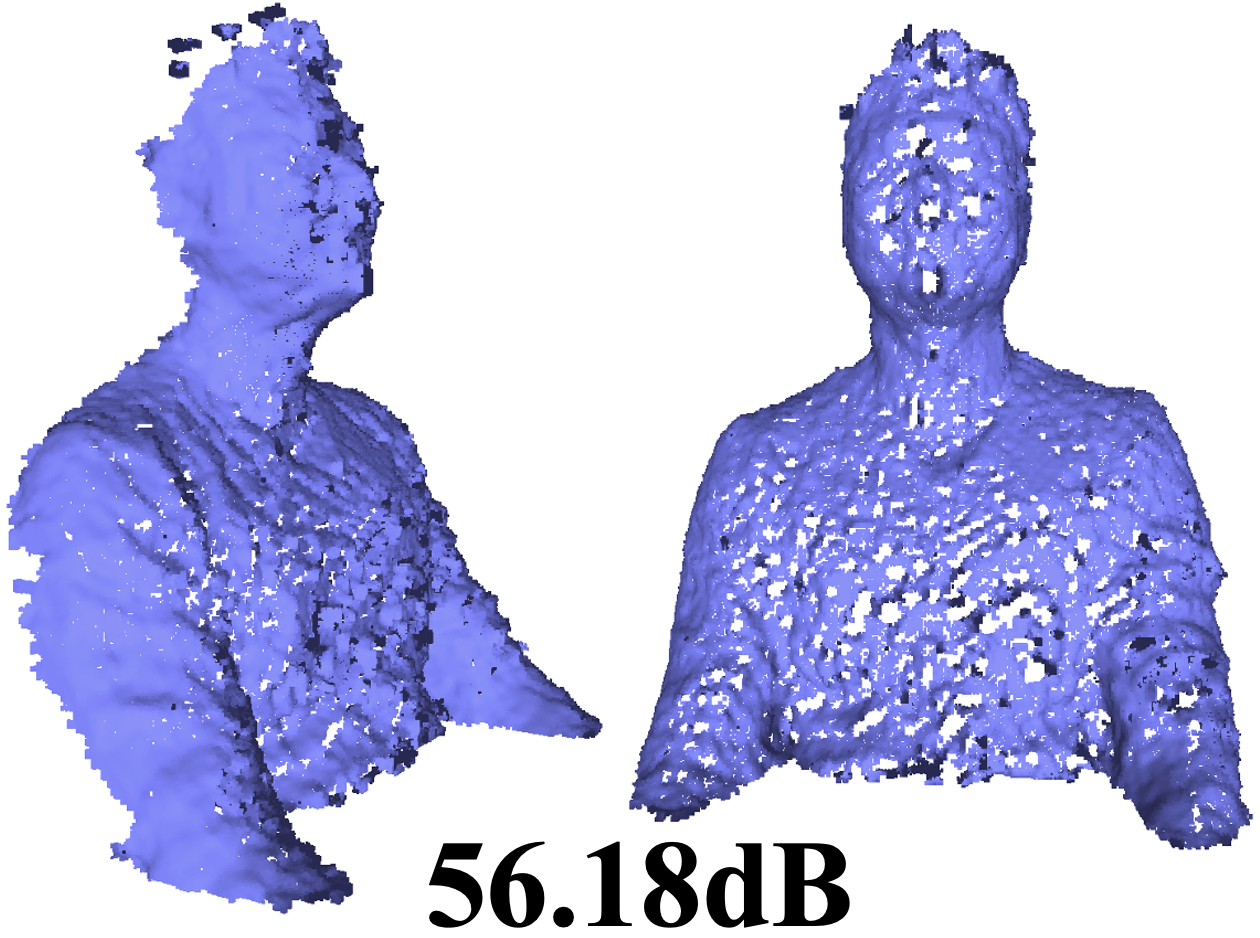}}
\subfigure[VVRec-C]{
\includegraphics[width=0.151\textwidth]{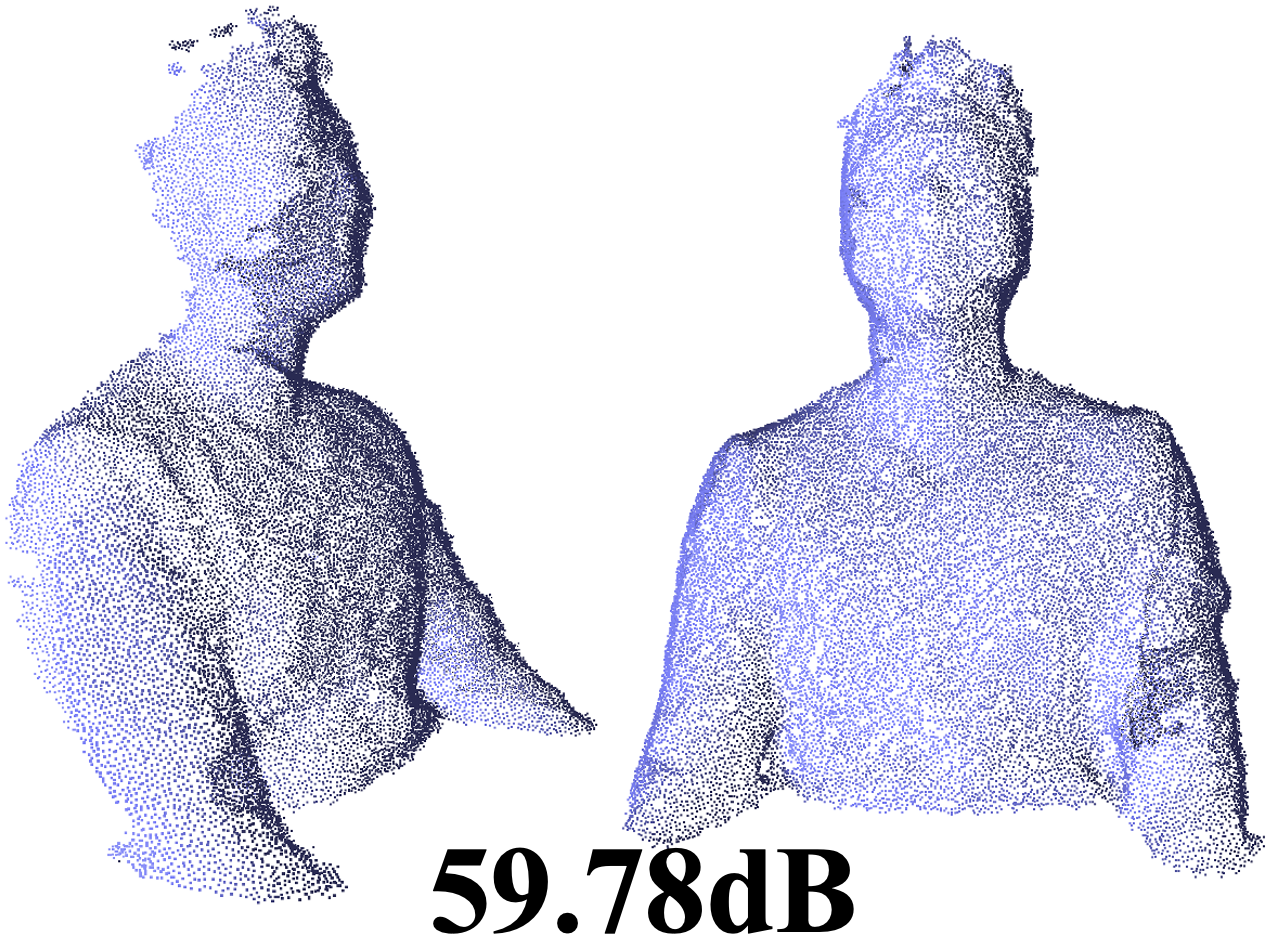}}
\subfigure[VVRec]{
\includegraphics[width=0.151\textwidth]{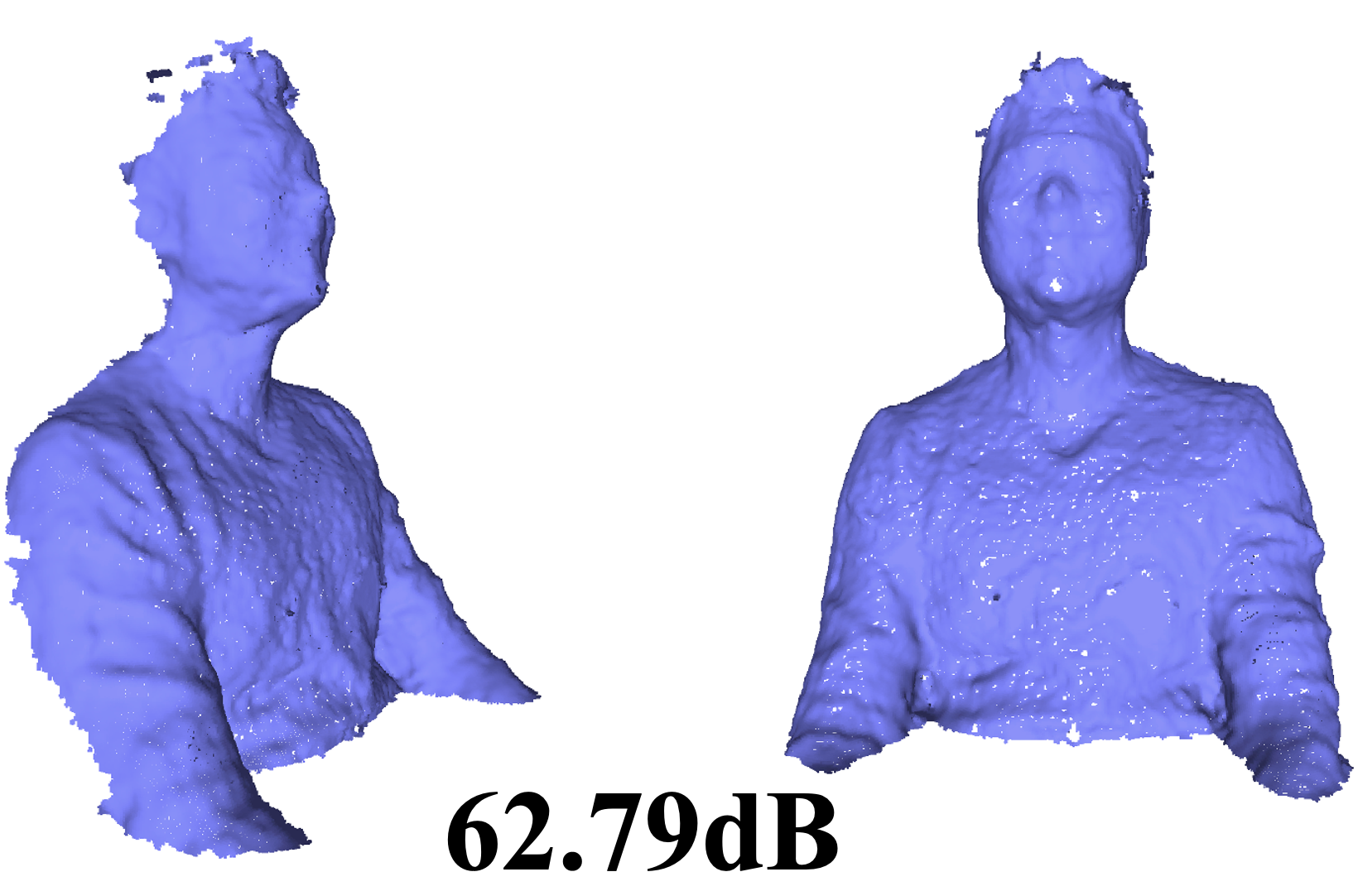}}
\caption{Visualization of ablation study on breakdown version of VVRec, measured by p2plane-PSNR.}
\label{fig_ablation}
\end{minipage}
\end{figure*}

\begin{figure}[t]
    \centering
    \begin{minipage}[t]{0.52\linewidth}
    \includegraphics[width=\textwidth]{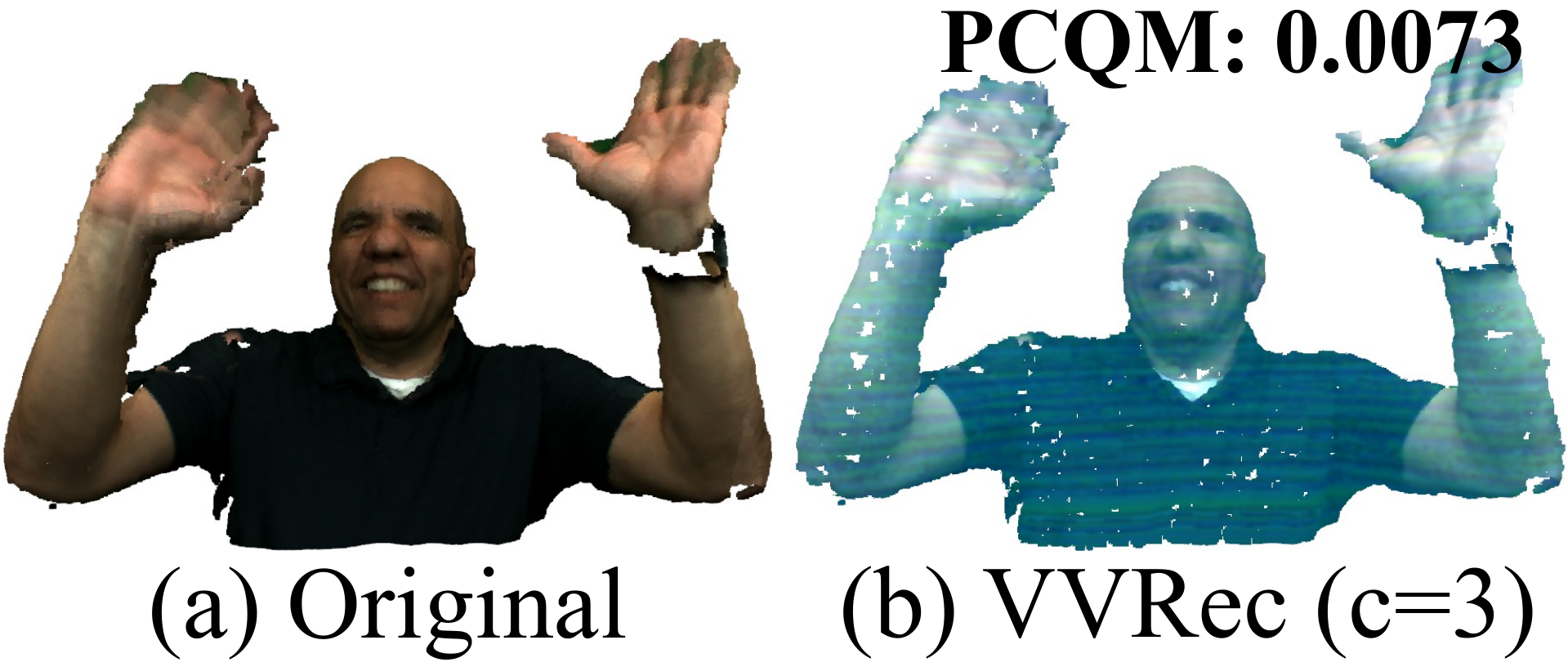}
    \caption{Study of color recovery on \textit{Ricardo@MVUB}.}
    \label{fig_color}
    \end{minipage}
    \begin{minipage}[t]{0.47\linewidth}
    \includegraphics[width=\textwidth]{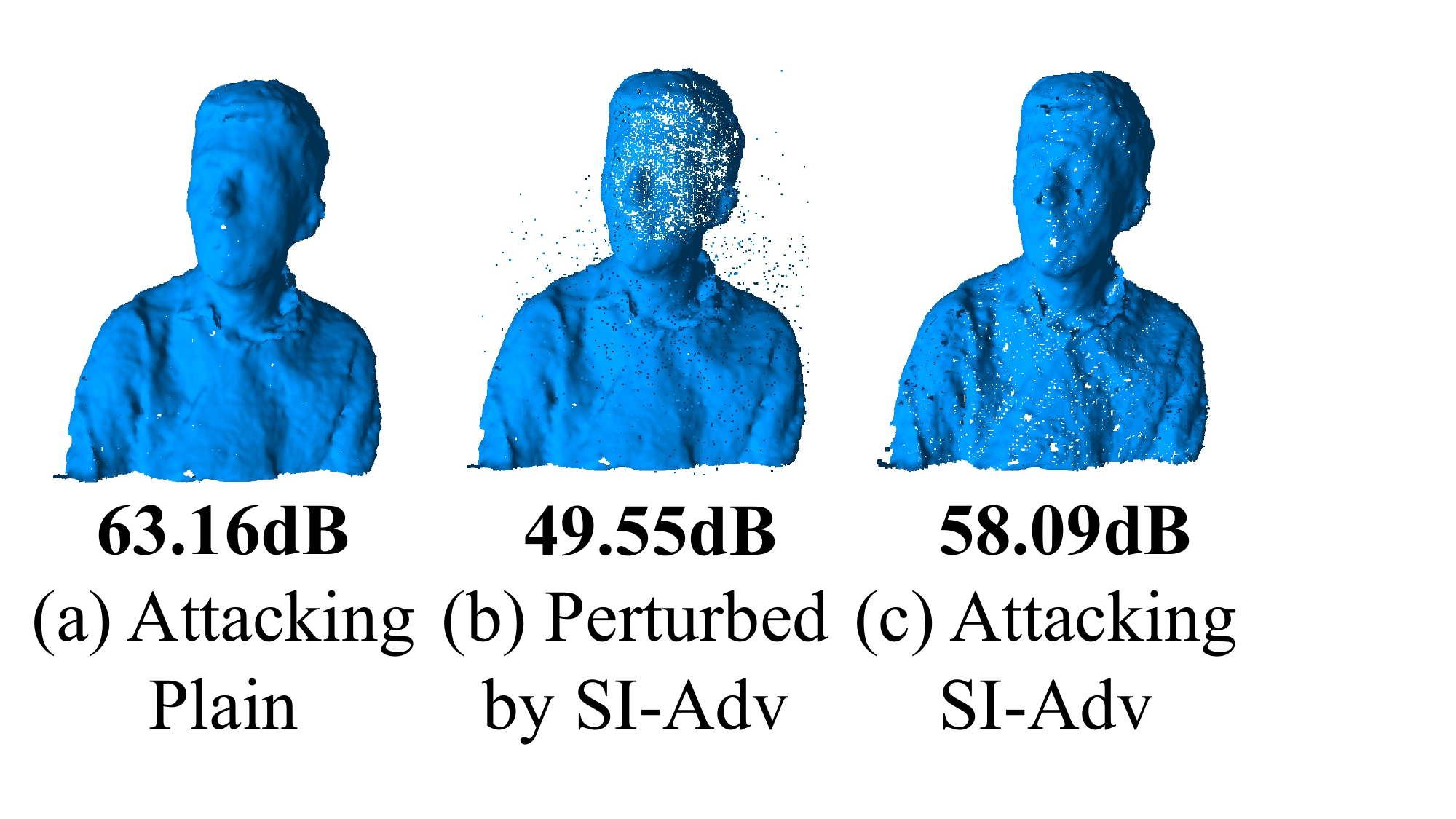}
    \caption{Study under protection on \textit{Andrew@MVUB}.}
    \label{fig_protect}
    \end{minipage}
\end{figure}

\subsection{Reconstruction Performance}
The performance of VVRec is measured by M1 and M2 between the reconstructed point cloud and the original one. In this context, the attacker's dataset is used for training the attacker model, while the testing dataset is the dataset encoded by the victim encoder. 

As shown in Tab.~\ref{tab_overall}, when the attacker dataset is MVUB and the victim dataset is 8iv2, VVRec exhibits only a marginal decline at about 18.91\% in M1, averaging on seven victim models compared to the Optimal, representing minimal quality degradation among all attacking schemes. The supervised method records the lowest M1 and M2, likely due to its training data being utterly different from the victim model. Additionally, VVRec outperforms SP-GAN with an improvement of M1 up to 33.18\%. The results show that the latent diffusion scheme in VVRec can converge to the distribution of the decoder model, leading to high reconstruction performance.
Notably, SP-GAN shows varying performance under different victim models, possibly attributed to challenges in convergence and local optimum traps.
We observe consistent results for other combinations of the attacker and victim datasets. VVRec exhibits up to 37.20\% improvement in M1 compared to other methods, with only 12.02\% to 31.28\% lower performance than the Optimal. Moreover, compared to the case when the attacker dataset is MVUB, the reconstruction performance of all attack methods is enhanced when trained on Owlii and UVG-VPC. This improvement can be attributed to the similarity in volumetric video content shared by Owlii, UVG-VPC and 8iv2, all featuring full bodies, while MVUB represents half-upper body content. The improvement is more apparent when the attacker training dataset is a subset of 8iv2. Fig.~\ref{fig_overall} displays the visual comparison between VVRec and baselines vividly.

\subsection{Ablation Study}
To explore the virtue of each component in VVRec, we utilize three breakdown versions of VVRec, \textit{VVRec-A}, \textit{VVRec-B}, and \textit{VVRec-C}. \textit{VVRec-A} does not have SLDM and LPDGM; only LPG and RD are implemented. \textit{VVRec-B} replaces the Gamma distribution in LPDGM with the Gaussian. \textit{VVRec-C} does not apply refinement in VVRec.

We use M2 as the assessment metric. We train attacker models on MVUB and attack the intermediate results generated by PCGCv2 trained on 8iv2.
In Fig.~\ref{fig_ablation}, VVRec-A exhibits inferior performance compared to VVRec, achieving the lowest reconstruction quality due to the absence of the diffusion model scheme, making VVRec-C theoretically a supervised method suffering from serious data distribution shift. 
VVRec-B lags behind VVRec by 6.61dB, indicating that the Gaussian distribution in high-dimensional latent points performs worse than the Gamma distribution in the latent diffusion model. 
VVRec-C, lacking a refinement module, experiences a 4.79\% reduction in quality compared to VVRec, underscoring the significance of the refinement module in further enhancing reconstruction performance.

\subsection{Study On Color Recovery}
Here, we try to investigate VVRec's capability of color recovery.
As mentioned in Section~\ref{sec:design-lpg}, VVRec supports color representations when appropriate $c$ values are assigned. We set LVAC \cite{lvac} as the victim model as it can encode the color representation within intermediate results. We use PCQM~\cite{meynet2020pcqm} to evaluate those point clouds with color, and the visualization of attacking results of trace \textit{Ricardo} from MVUB and the one from victim decoder is shown in Fig.~\ref{fig_color}. Compared to the Optimal, the overall average PCQM degradation is only 26.13\%, indicating that VVRec can reconstruct both the geometry and color information of point clouds in high quality.

\subsection{Study On Possible Protection}
Since no existing method is specifically designed for protecting against point cloud reconstruction attacks, we explore a typical protection solution, SI-Adv~\cite{huang2022shape}, designed for attribute inference attacks protection. It introduces adversarial perturbations by injecting a few points into the point clouds to confuse the analytics results. For instance, it adds noise points on the face to disturb the gender classification. The visualization results are displayed in Fig.~\ref{fig_protect}. The overall average reconstruction accuracy, measured by p2plane-PSNR, indicates that this protection approach has only a minimal impact on reconstruction performance, with less than 10\% degradation.
\section{Limitations}
Regarding limitations, VVRec is designed specifically for DL-based volumetric upstreaming, where privacy is a concern during transmission. It does not apply to downstream applications, as their content is meant for public dissemination.
Secondly, we assume that the attacker has no latency constraints, as even recovering a single frame could result in significant privacy breaches. However, this assumption may be limiting in scenarios that require rapid response due to sampling delays.
Third, as protection solutions for DL-based streaming continue to evolve, our attack schemes could potentially be countered by emerging defensive approaches.

\section{Conclusions}
In conclusion, this paper presents VVRec, a novel reconstruction attack scheme targeting DL-based volumetric video upstreaming. 
VVRec demonstrates the ability to reconstruct high-quality point clouds from intercepted intermediate results using four well-trained neural network models: two latent diffusion models (SLDM, LPGDM), a generator (LPG), and a decoder (RD). Leveraging the latest latent diffusion models with Gamma distribution and a refinement module, VVRec excels in reconstruction quality and color recovery and surpasses existing defenses. The significance of our work extends to understanding the privacy threats in volumetric video applications where safeguarding sensitive information during streaming is crucial.

\section*{Acknowledgments}
Dan Wang's work is supported by RGC GRF 15200321, 15201322, 15230624, RGC-CRF C5018-20G, ITC ITF-ITS/056/22MX, and PolyU 1-CDKK, G-SAC8.

\bigskip

\bibliography{aaai25}
\appendix
\section{Appendix}
\subsection{VAE Models}
Our latent point generator (LPG) and reconstruction decoder (RD) are constructed as VAEs. We display the architecture of LPG and RD in Tab.~\ref{tab_appendix1} and Tab.~\ref{tab_appendix2}, respectively. All dropout layers in the VAEs are implemented with a dropout probability of 0.1. To fulfill conditioning on the shape latent, we replace them with the AdaGN layers. The initial weight scale is 0.1. We also add residual connections in various layers. In the LPG, the raw input point cloud plus the product of 0.01 and the mean of the predicted latent point form the final output. In the RD, the final output is the summation of the sampled latent point cloud and 0.01 times the predicted point cloud.

\begin{table}[h]
\centering
\caption{Architecture of LPG.}
\resizebox{\linewidth}{!}{
\setlength{\tabcolsep}{2pt}{
\begin{tabular}{ccccc}
\hline
\multicolumn{5}{c}{Input: shape latent $z_0$} \\ \hline
\multicolumn{5}{c}{Output: latent points $h_0$} \\ \hline
\hline
 & SA1 & SA2 & SA3 & SA4 \\ \hline
 \# PVC Layers & 2 & 1 & 1 & N/A \\
 \# PVC hidden dimension & 32 & 64 & 128 & N/A \\
 \# PVC voxel grid size & 32 &16 &8 &N/A \\
\# Grouper center &1024 &256 &64 &16 \\
\# Grouper radius &0.1 &0.2 &0.4 &0.8 \\
\# Grouper neighbors &32 &32 &32 &32 \\
\# MLP layers &2 &2 &2 &3 \\
\# MLP output dimension &32,32 &64,128 &128,256 &128,128,128 \\
\# Attention dimension &N/A &128 &N/A &N/A \\ \hline
\hline
\multicolumn{5}{c}{Global attention, hidden dimension of 256} \\ \hline
\hline
 & FP1 & FP2 & FP3 & FP4 \\ \hline
\# MLP layers &2 &2 &2 &3 \\
\# MLP output dimension &128,128 &128,128 &128,128 &128,128,64 \\
\# PVC layers &3 &3 &2 &2 \\
\# PVC hidden dimension &128 &128 &128 &64 \\
\# PVC voxel grid size &8 &8 &16 &32 \\
\# Attention dimension &N/A &128 &N/A &N/A \\ \hline
\hline
\multicolumn{5}{c}{MLP: (64,128)} \\
\multicolumn{5}{c}{Dropout} \\
\multicolumn{5}{c}{Linear: (128, $2\times(3+c+D_h)$)} \\
\hline
\end{tabular}
\label{tab_appendix1}
}
}
\end{table}

\begin{table}[h!]
\centering
\caption{Architecture of RD.}
\resizebox{\linewidth}{!}{
\setlength{\tabcolsep}{2pt}{
\begin{tabular}{ccccc}
\hline
\multicolumn{5}{c}{Input: shape latent $z_0$, latent points $h_0$} \\ \hline
\multicolumn{5}{c}{Output: reconstructed point cloud $x'$} \\ \hline
\hline
 & SA1 & SA2 & SA3 & SA4 \\ \hline
 \# PVC Layers & 2 & 1 & 1 & N/A \\
 \# PVC hidden dimension & 32 & 64 & 128 & N/A \\
 \# PVC voxel grid size & 32 &16 &8 &N/A \\
\# Grouper center &1024 &256 &64 &16 \\
\# Grouper radius &0.1 &0.2 &0.4 &0.8 \\
\# Grouper neighbors &32 &32 &32 &32 \\
\# MLP layers &2 &2 &2 &3 \\
\# MLP output dimension &32,32 &64,128 &128,256 &128,128,128 \\
\# Attention dimension &N/A &128 &N/A &N/A \\ \hline
\hline
\multicolumn{5}{c}{Global attention, hidden dimension of 256} \\ \hline
\hline
 & FP1 & FP2 & FP3 & FP4 \\ \hline
\# MLP layers &2 &2 &2 &3 \\
\# MLP output dimension &128,128 &128,128 &128,128 &128,128,64 \\
\# PVC layers &3 &3 &2 &2 \\
\# PVC hidden dimension &128 &128 &128 &64 \\
\# PVC voxel grid size &8 &8 &16 &32 \\ \hline
\hline
\multicolumn{5}{c}{MLP: (64,128)} \\
\multicolumn{5}{c}{Dropout} \\
\multicolumn{5}{c}{Linear: (128, $3+c$)} \\
\hline
\end{tabular}
\label{tab_appendix2}
}
}
\end{table}

\subsection{Additional Results}
In this section, we display the additional results of the reconstruction performance of VVRec and other baselines. In Tab.~\ref{tab_lossless}, we show the reconstruction quality of VVRec and baselines targeting lossless victim models. Different from lossy victim models, we cannot tune the bit rate of pre-trained lossless victim models. Therefore, all results are adopted in the case that all lossless victim models encode the point cloud at the lowest bit rate they can achieve. Similar to the case of lossy victim models, VVRec outperforms all other baselines significantly on lossless victim models. Specifically, even for the strongest baseline (SP-GAN), VVRec still achieves up to 33.18\% improvement in M1, demonstrating that VVRec generalizes well on lossless victim models. In Tab.~1 of our paper, we have displayed the reconstruction quality of VVRec and baselines when the bit rate of lossy victim models is set to 0.4 bpp. In this section, we will further demonstrate the reconstruction quality of VVRec and baselines under a bit rate of 0.2 bpp and 0.1 bpp in Tab.~\ref{tab_0.4} and Tab.~\ref{tab_0.2}, respectively. When the bit rate of victim models drops, the upper bound of reconstruction quality (\textit{"Optimal"}) and reconstruction quality of all attack methods, including VVRec, also decreases. This is because a lower bit rate introduces more noise into the encoding-decoding process. But VVRec still displays its superiority over other baselines. According to Tab.~\ref{tab_0.4}, when the attacker dataset is MVUB and the victim dataset is 8iv2, VVRec still shows 16.90\%-37.41\% improvement over all other baselines in M1. And we can observe similar results when the bit rate is 0.1 bpp.

\begin{table}[h!]
\caption{Reconstruction quality of VVRec and baselines measured by p2p-PSNR (M1$\uparrow$) and p2plane-PSNR (M2$\uparrow$) on lossless victim models. }
\centering
\resizebox{\linewidth}{!}{
\setlength{\tabcolsep}{3pt}{
\footnotesize
\begin{tabular}{c|c|l|cc|cc}
\hline
\multirow{2}{*}{\textbf{Attacker Dataset}} & \multirow{2}{*}{\textbf{Victim Dataset}} & \textbf{Victim Encoder} & \multicolumn{2}{c|}{\textbf{\textit{VoxelDNN}}} & \multicolumn{2}{c}{\textbf{\textit{MSVoxelDNN}}} \\ \cline{3-7}
& & \textbf{Metrics} & M1 & M2 & M1 & M2 \\ \hline
\multirow{5}{*}{MVUB} & \multirow{5}{*}{8iv2} & \textit{Supervised} & 39.38 & 44.16 & 37.26  & 43.09  \\ 
& &  \textit{SP-GAN} & 44.08 & 49.16 & 42.26 &47.07  \\
& &  \textit{VG-VAE}  & 42.04 & 47.28 & 40.26 &45.27 \\
& &  \textit{PD-Flow}  & 41.05 & 46.16 & 39.14 &44.02  \\
& &  \textit{VVRec}  & \textbf{56.79} & \textbf{61.06} & \textbf{56.28} & \textbf{61.26} \\ \cline{1-7}
\multirow{5}{*}{MVUB} & \multirow{5}{*}{Owlii} & \textit{Supervised} & 39.72 & 44.01 & 37.07  & 42.74   \\ 
& & \textit{SP-GAN} & 44.09 & 49.05 & 42.03 & 47.11  \\
& & \textit{VG-VAE}  & 42.12 & 47.19 & 39.98 & 44.96 \\
& &\textit{PD-Flow}  & 40.85 & 45.93 & 38.98 & 43.87  \\
& & \textit{VVRec} & \textbf{56.72} & \textbf{60.89} & \textbf{55.88} & \textbf{60.32}  \\ \cline{1-7}
\multirow{5}{*}{MVUB} & \multirow{5}{*}{UVG-VPC} & \textit{Supervised} & 39.01 & 43.56 & 36.67  & 42.59  \\ 
& & \textit{SP-GAN} & 43.54 & 48.77 & 41.53 & 46.47 \\
& & \textit{VG-VAE}  & 41.38 & 46.65 & 39.71 & 44.96  \\
& &\textit{PD-Flow}  & 40.86 & 45.74 & 38.65 & 43.62  \\
& & \textit{VVRec} & \textbf{56.28} & \textbf{60.32} & \textbf{55.61} & \textbf{59.88}  \\ \cline{1-7}
\multirow{5}{*}{Owlii} & \multirow{5}{*}{8iv2} & \textit{Supervised}  & 44.27 & 49.15 & 42.35 &47.07 \\
& & \textit{SP-GAN} & 47.05 & 52.85 & 44.36 &49.57 \\
& & \textit{VG-VAE}& 45.25 & 50.86 & 42.33 &47.58  \\
& & \textit{PD-FLow} & 44.01 & 49.56 & 41.25 &46.55  \\
& & \textit{VVRec} & \textbf{58.23} & \textbf{61.35} & \textbf{57.89} & \textbf{61.99} \\ \cline{1-7}
 \multirow{5}{*}{UVG-VPC} & \multirow{5}{*}{8iv2} & \textit{Supervised} & 44.02 & 49.11 & 42.36 &47.08 \\
& & \textit{SP-GAN} & 47.24 & 52.97 & 44.08 &49.18  \\
& & \textit{VG-VAE} & 45.10 & 50.76 & 42.35 &47.54 \\
& & \textit{PD-FLow} & 43.79 & 49.17 & 41.05 &46.29  \\
& & \textit{VVRec} & \textbf{58.02} & \textbf{61.08} & \textbf{57.29} & \textbf{61.64} \\ \cline{1-7}

\end{tabular}
\label{tab_lossless}
}
}
\end{table}

\begin{table*}[!t]
\caption{Reconstruction quality of VVRec and baselines measured by p2p-PSNR (M1$\uparrow$) and p2plane-PSNR (M2$\uparrow$) with bitrate 0.2bpp. }
\centering
\resizebox{0.85\linewidth}{!}{
\setlength{\tabcolsep}{3pt}{
\footnotesize
\begin{tabular}{c|c|l|cc|cc|cc|cc|cc}
\hline
\multirow{2}{*}{\textbf{Attacker Dataset}} & \multirow{2}{*}{\textbf{Victim Dataset}} & \textbf{Victim Encoder} & \multicolumn{2}{c|}{\textbf{\textit{PCGCv1}}} & \multicolumn{2}{c|}{\textbf{\textit{PCGCv2}}} & \multicolumn{2}{c|}{\textbf{\textit{SparsePCGC}}} & \multicolumn{2}{c|}{\textbf{\textit{Pcc-geo-cnn-v1}}} & \multicolumn{2}{c}
{\textbf{\textit{Pcc-geo-cnn-v2}}} \\ \cline{3-13}
& & \textbf{Metrics} & M1 & M2 & M1 & M2& M1 & M2& M1 & M2 & M1 & M2 \\ \hline
\multirow{6}{*}{MVUB} & \multirow{6}{*}{8iv2} & \textcircled{1} \textit{Optimal} & 68.27 & 72.93 & 70.14 & 73.96 & 73.04 & 77.16 & 64.75 & 68.47 & 67.17 & 70.89 \\  \cline{3-13}
& & \textcircled{2} \textit{Supervised} & 43.38 & 47.51 & 39.18 & 43.51 & 38.45 & 42.94 & 44.38 & 49.05 & 39.62 & 43.71 \\ 
& & \textcircled{3} \textit{SP-GAN} & 47.69 & 52.09 & 43.52 & 47.69 & 42.28 & 46.51 & 45.98 & 50.34 & 43.52 & 48.17 \\
& &  \textcircled{4} \textit{VG-VAE}  & 45.79 & 50.09 & 41.39 & 45.50 & 40.05 & 44.14 & 43.90 & 48.29 & 41.44 & 46.75 \\
& &  \textcircled{5} \textit{PD-Flow}  & 45.12 & 49.16 & 39.83 & 44.67 & 39.22 & 43.72 & 42.49 & 46.77 & 40.44 & 45.46 \\
& & \textcircled{6} \textit{VVRec} & \textbf{55.75} & \textbf{60.53} & \textbf{53.29} & \textbf{57.78} & \textbf{50.93} & \textbf{54.99} & \textbf{56.28} & \textbf{60.46} & \textbf{54.44} & \textbf{59.04} \\ \cline{1-13}
\multirow{6}{*}{MVUB} & \multirow{6}{*}{Owlii} & \textit{Optimal} & 68.72 & 72.70 & 69.88 & 73.76 & 73.17 & 76.86 & 64.73 & 68.66 & 66.93 & 70.33  \\  \cline{3-13}
& & \textit{Supervised} & 43.35 & 47.52 & 39.10 & 43.44 & 38.40 & 42.76 & 44.68 & 48.73 & 39.58 & 43.86 \\ 
& & \textit{SP-GAN} & 47.51 & 52.34 & 43.60 & 47.49 & 42.20 & 46.63 & 45.82 & 50.43 & 43.47 & 48.50\\
& & \textit{VG-VAE}  & 45.64 & 50.08 & 41.32 & 45.42 & 40.10 & 44.16 & 43.99 & 48.20 & 41.65 & 46.72 \\
& &\textit{PD-Flow}  & 44.97 & 49.13 & 39.77 & 44.57 & 39.29 & 43.89 & 42.57 & 46.80 & 40.73 & 45.52 \\
& & \textit{VVRec} & \textbf{55.78} & \textbf{60.44} & \textbf{53.30} & \textbf{57.88} & \textbf{51.17} & \textbf{55.02} & \textbf{56.54} & \textbf{60.44} & \textbf{54.40} & \textbf{59.04} \\ \cline{1-13}
\multirow{6}{*}{MVUB} & \multirow{6}{*}{UVG-VPC} & \textit{Optimal} & 69.50 & 73.58 & 71.06 & 74.46 & 73.49 & 77.89 & 65.66 & 69.51 & 68.09 & 71.16  \\  \cline{3-13}
& & \textit{Supervised} & 42.94 & 46.76 & 38.62 & 42.94 & 37.89 & 42.12 & 43.77 & 48.39 & 38.97 & 43.57\\ 
& & \textit{SP-GAN} & 47.03 & 51.34 & 42.74 & 47.54 & 41.74 & 45.84 & 45.31 & 49.66 & 42.67 & 47.78 \\
& & \textit{VG-VAE}  & 45.18 & 49.74 & 40.83 & 45.00 & 39.69 & 43.86 & 43.25 & 47.79 & 41.06 & 45.67 \\
& &\textit{PD-Flow}  & 44.75 & 48.60 & 39.03 & 44.01 & 38.53 & 43.11 & 41.87 & 46.49 & 39.91 & 44.86\\
& & \textit{VVRec} & \textbf{55.32} & \textbf{60.01} & \textbf{52.93} & \textbf{57.14} & \textbf{50.66} & \textbf{54.89} & \textbf{55.84} & \textbf{60.16} & \textbf{53.72} & \textbf{58.59} \\ \cline{1-13}
\multirow{6}{*}{Owlii} & \multirow{6}{*}{8iv2} & \textit{Optimal} & 68.71 & 72.85 & 70.06 & 73.81 & 73.42 & 77.25 & 65.02 & 68.39 & 67.17 & 71.04  \\  \cline{3-13}
& & \textit{Supervised} & 47.68 & 51.98 & 43.59 & 47.89 & 42.14 & 46.64 & 46.75 & 50.85 & 44.23 & 48.70 \\
& & \textit{SP-GAN} & 49.68 & 53.71 & 46.47 & 51.14 & 45.43 & 50.20 & 48.60 & 52.83 & 46.48 & 51.09 \\
& & \textit{VG-VAE} & 47.60 & 51.77 & 44.49 & 49.23 & 43.70 & 48.15 & 46.63 & 50.85 & 44.67 & 49.33  \\
& & \textit{PD-FLow} & 46.88 & 50.41 & 42.62 & 48.30 & 42.48 & 46.91 & 45.84 & 49.74 & 42.95 & 47.32 \\
& & \textit{VVRec} & \textbf{57.21} & \textbf{61.79} & \textbf{55.40} & \textbf{59.37} & \textbf{53.09} & \textbf{58.10} & \textbf{56.85} & \textbf{61.34} & \textbf{55.93} & \textbf{60.04} \\ \cline{1-13}
 \multirow{6}{*}{UVG-VPC} & \multirow{6}{*}{8iv2} & \textit{Optimal} & 68.35 & 73.31 & 70.14 & 74.12 & 73.19 & 77.33 & 64.96 & 68.32 & 67.38 & 70.60 \\  \cline{3-13}
& & \textit{Supervised} & 47.68 & 51.56 & 43.44 & 48.20 & 41.77 & 46.63 & 46.24 & 50.27 & 44.08 & 48.86 \\
& & \textit{SP-GAN} &49.87 & 53.46 & 45.73 & 50.63 & 45.57 & 50.07 & 48.61 & 52.78 & 46.48 & 51.06 \\
& & \textit{VG-VAE} &47.73 & 51.67 & 44.57 & 49.35 & 42.73 & 47.65 & 48.95 & 53.37 & 44.68 & 49.24 \\
& & \textit{PD-FLow} & 46.52 & 50.70 & 42.72 & 48.21 & 42.30 & 46.90 & 45.51 & 49.71 & 42.78 & 47.48\\
& & \textit{VVRec} & \textbf{57.62} & \textbf{61.80} & \textbf{55.05} & \textbf{59.31} & \textbf{52.96} & \textbf{57.83} & \textbf{57.30} & \textbf{62.33} & \textbf{55.59} & \textbf{60.58} \\ \cline{1-13}
\end{tabular}
\label{tab_0.4}
}
}
\end{table*}

\begin{table*}[!h]
\caption{Reconstruction quality of VVRec and baselines measured by p2p-PSNR (M1$\uparrow$) and p2plane-PSNR (M2$\uparrow$) with bitrate 0.1bpp. }
\centering
\resizebox{0.85\linewidth}{!}{
\setlength{\tabcolsep}{3pt}{
\footnotesize
\begin{tabular}{c|c|l|cc|cc|cc|cc|cc}
\hline
\multirow{2}{*}{\textbf{Attacker Dataset}} & \multirow{2}{*}{\textbf{Victim Dataset}} & \textbf{Victim Encoder} & \multicolumn{2}{c|}{\textbf{\textit{PCGCv1}}} & \multicolumn{2}{c|}{\textbf{\textit{PCGCv2}}} & \multicolumn{2}{c|}{\textbf{\textit{SparsePCGC}}} & \multicolumn{2}{c|}{\textbf{\textit{Pcc-geo-cnn-v1}}} & \multicolumn{2}{c}
{\textbf{\textit{Pcc-geo-cnn-v2}}} \\ \cline{3-13}
& & \textbf{Metrics} & M1 & M2 & M1 & M2& M1 & M2& M1 & M2 & M1 & M2 \\ \hline
\multirow{6}{*}{MVUB} & \multirow{6}{*}{8iv2} & \textcircled{1} \textit{Optimal} & 64.53 & 69.47 & 66.08 & 70.16 & 68.81 & 72.39 & 61.14 & 64.65 & 63.77 & 66.58  \\  \cline{3-13}
& & \textcircled{2} \textit{Supervised} & 40.97 & 44.66 & 36.76 & 41.04 & 36.35 & 40.69 & 42.13 & 46.33 & 37.29 & 41.36 \\ 
& & \textcircled{3} \textit{SP-GAN} & 45.24 & 49.40 & 40.86 & 45.19 & 39.70 & 44.17 & 43.24 & 47.48 & 41.24 & 45.93 \\
& &  \textcircled{4} \textit{VG-VAE}  & 43.25 & 47.55 & 39.12 & 43.30 & 37.78 & 41.67 & 41.60 & 45.30 & 39.35 & 43.91 \\
& &  \textcircled{5} \textit{PD-Flow}  & 42.57 & 46.33 & 37.75 & 42.37 & 36.99 & 41.51 & 40.39 & 44.18 & 38.27 & 43.02 \\
& & \textcircled{6} \textit{VVRec} & \textbf{53.05} & \textbf{56.91} & \textbf{50.53} & \textbf{54.31} & \textbf{48.40} & \textbf{52.09} & \textbf{52.96} & \textbf{57.72} & \textbf{50.97} & \textbf{55.81} \\ \cline{1-13}
\multirow{6}{*}{MVUB} & \multirow{6}{*}{Owlii} & \textit{Optimal} & 64.53 & 69.11 & 65.62 & 69.34 & 68.95 & 72.75 & 61.27 & 65.20 & 63.62 & 66.48 \\  \cline{3-13}
& & \textit{Supervised} & 41.29 & 45.07 & 36.76 & 40.84 & 36.26 & 40.82 & 42.05 & 45.90 & 37.59 & 41.56 \\ 
& & \textit{SP-GAN} & 45.10 & 49.65 & 40.81 & 45.28 & 39.98 & 44.13 & 43.51 & 47.36 & 41.14 & 45.71 \\
& & \textit{VG-VAE}  & 43.28 & 47.44 & 39.23 & 43.31 & 38.00 & 42.06 & 41.45 & 45.56 & 39.47 & 44.22 \\
& &\textit{PD-Flow}  & 42.56 & 45.94 & 37.77 & 42.12 & 36.90 & 41.50 & 40.25 & 44.30 & 38.38 & 42.80 \\
& & \textit{VVRec} & \textbf{53.08} & \textbf{57.39} & \textbf{50.60} & \textbf{54.54} & \textbf{48.27} & \textbf{52.06} & \textbf{53.40} & \textbf{56.93} & \textbf{51.03} & \textbf{56.25} \\ \cline{1-13}
\multirow{6}{*}{MVUB} & \multirow{6}{*}{UVG-VPC} & \textit{Optimal} & 65.77 & 69.55 & 66.88 & 69.84 & 69.47 & 72.98 & 62.55 & 66.00 & 64.31 & 67.41  \\  \cline{3-13}
& & \textit{Supervised} & 40.64 & 44.01 & 36.55 & 40.51 & 35.59 & 39.86 & 41.52 & 45.89 & 36.71 & 40.92 \\ 
& & \textit{SP-GAN} & 44.41 & 48.58 & 40.49 & 44.60 & 39.19 & 43.33 & 42.52 & 47.31 & 40.69 & 45.18 \\
& & \textit{VG-VAE}  & 42.76 & 46.51 & 38.50 & 42.55 & 37.19 & 41.05 & 40.93 & 44.83 & 38.56 & 43.30 \\
& &\textit{PD-Flow}  & 42.36 & 45.70 & 36.94 & 41.56 & 36.30 & 40.31 & 39.54 & 43.71 & 37.93 & 42.17 \\
& & \textit{VVRec} & \textbf{52.01} & \textbf{56.54} & \textbf{50.10} & \textbf{53.71} & \textbf{47.63} & \textbf{51.85} & \textbf{53.20} & \textbf{56.93} & \textbf{50.79} & \textbf{55.81} \\ \cline{1-13}
\multirow{6}{*}{Owlii} & \multirow{6}{*}{8iv2} & \textit{Optimal} & 64.46 & 69.40 & 66.23 & 69.46 & 69.12 & 72.31 & 61.07 & 65.08 & 63.21 & 67.10  \\  \cline{3-13}
& & \textit{Supervised} & 45.18 & 48.88 & 40.89 & 45.67 & 39.75 & 44.13 & 44.05 & 48.49 & 41.76 & 46.34 \\
& & \textit{SP-GAN} & 46.60 & 50.78 & 44.07 & 48.35 & 43.09 & 47.00 & 45.88 & 50.10 & 44.04 & 48.25 \\
& & \textit{VG-VAE} & 44.94 & 49.21 & 42.10 & 46.38 & 41.23 & 45.68 & 44.27 & 47.65 & 42.13 & 46.58 \\
& & \textit{PD-FLow} & 44.47 & 47.49 & 40.24 & 45.52 & 39.85 & 44.28 & 43.04 & 46.91 & 40.56 & 44.82 \\
& & \textit{VVRec} & \textbf{53.95} & \textbf{57.97} & \textbf{51.80} & \textbf{55.87} & \textbf{49.90} & \textbf{54.69} & \textbf{53.35} & \textbf{57.97} & \textbf{53.22} & \textbf{57.01} \\ \cline{1-13}
 \multirow{6}{*}{UVG-VPC} & \multirow{6}{*}{8iv2} & \textit{Optimal} & 64.53 & 69.55 & 65.86 & 69.93 & 69.12 & 73.03 & 61.07 & 64.57 & 63.77 & 67.10  \\  \cline{3-13}
& & \textit{Supervised} &44.74 & 48.46 & 41.15 & 45.42 & 39.70 & 44.13 & 43.71 & 47.67 & 41.76 & 46.35\\
& & \textit{SP-GAN} &47.20 & 50.38 & 43.38 & 47.40 & 43.46 & 47.60 & 45.86 & 49.83 & 43.85 & 48.05 \\
& & \textit{VG-VAE} & 45.03 & 48.69 & 41.72 & 46.25 & 40.48 & 45.23 & 45.97 & 50.58 & 42.28 & 46.49 \\
& & \textit{PD-FLow} & 43.82 & 48.04 & 40.20 & 45.62 & 39.60 & 44.38 & 43.21 & 47.14 & 40.57 & 44.68 \\
& & \textit{VVRec} & \textbf{54.48} & \textbf{58.17} & \textbf{51.81} & \textbf{55.95} & \textbf{50.40} & \textbf{54.60} & \textbf{55.93} & \textbf{58.61} & \textbf{52.54} & \textbf{57.02} \\ \cline{1-13}
\end{tabular}
\label{tab_0.2}
}
}
\end{table*}

\end{document}